\documentclass[aps,pre,twocolumn,superscriptaddress,amsmath,amssymb,nofootinbib]{revtex4-1}
\usepackage{graphicx,enumitem}
\usepackage{multirow,dcolumn}
\usepackage{txfonts}
\usepackage{hyperref}
\usepackage{soul,cancel}
\usepackage{epsf,amsmath,amssymb,verbatim,color,multirow,pifont,graphicx,cleveref,tabularx}
\usepackage[dvipsnames]{xcolor}
\usepackage{rotating}
\usepackage{ulem}
\usepackage{amsmath,amsfonts,amssymb,bm,cancel}
\usepackage{kotex}

\newcommand{\cnew}[1]{\textcolor{cyan}{#1}}

\usepackage{tikz}

\newcommand{\avg}[1]{\langle{#1}\rangle}
\newcommand{\Avg}[1]{\left\langle{#1}\right\rangle}
\def\bea{\begin{eqnarray}}
\def\eea{\end{eqnarray}}

\def\Tr{\mathbf{T}}
\def\I{\textrm{I}}

\def\n{\textrm{n}}
\def\a{\textrm{a}}
\def\o{\textrm{o}}
\def\min{\textrm{min}}
\def\I{\textrm{I}}

\def\C{\mathcal{C}}
\def\S{\mathcal{S}}
\def\T{\mathcal{T}}
\def\B{\mathcal{B}}
\def\A{\mathcal{A}}

\begin{document}

\title{Link overlap influences  opinion dynamics on multiplex networks of Ashkin-Teller spins}
\author{Cook Hyun Kim}
\affiliation{CCSS, CTP and Department of Physics and Astronomy, Seoul National University, Seoul 08826, Korea}
\author{Minjae Jo}
\affiliation{CCSS, CTP and Department of Physics and Astronomy, Seoul National University, Seoul 08826, Korea}
\author{J. S. Lee}
\affiliation{School of Physics, Korea Institute for Advanced Study, Seoul 02455, Korea}
\author{G. Bianconi}
\affiliation{School of Mathematical Sciences, Queen Mary University of London, E1 4GF, London, United Kingdom}
\affiliation{Alan Turing Institute, The British Library, NW1 2DB, London, United Kingdom}
\author{B. Kahng}
\affiliation{Center for Complex Systems, KI of Grid Modernization, Korea Institute of Energy Technology, Naju, Jeonnam 58217, Korea}

\begin{abstract}
Consider a multiplex network formed by two layers indicating social interactions: the first layer is a friendship network and the  second layer is a  network of business relations. In this duplex network each pair of individuals can be connected  in different ways: they can be connected by a friendship  but not connected by a business relation, they can be connected  by a business relation without being friends, or they can be simultaneously friends and in a business relation. In the latter case we say that the links in different layers overlap.  These three types of connections are called multilinks and the multidegree indicates the sum of multilinks of a given type that are incident to a given node. 
Previous opinion models  on multilayer networks have mostly neglected the effect of link overlap.  Here we show that link overlap  can have important effects in  the formation of a majority opinion. Indeed, the formation of a majority opinion can be significantly influenced by the statistical properties  of  multilinks, and in particular by the  multidegree distribution.  To quantitatively address this problem, we study a simple spin model, called the Ashkin-Teller model including 2-body and 4-body interactions between nodes in different layers. Here we fully investigate the rich phase  diagram of this model which includes a large variety of phase transitions. Indeed the phase diagram or the model displays continuous, discontinuous, and hybrid phase transitions, and successive jumps of the order parameters within the Baxter phase. 
\end{abstract}

\maketitle

\section{INTRODUCTION}

Over the past two decades, network theory~\cite{NS,Doro_book,Newman_book,jkps,Doro_crit} has provided the pivotal framework for characterizing the interplay between graph structures and dynamics of complex systems.
Recently, multilayer networks~\cite{Bianconi2018multilayer,Buldyrev2010,PhysicsReports,Kivela,Goh_review} are attracting considerable scientific interest. These network of networks are able to integrate information on various types of links characterizing complex systems where interactions have different nature and connotation. Therefore, they provide a useful perspective for analyzing complex social, transportation, or biological systems~\cite{Thurner,Weighted,Boccaletti,Bullmore2009,Makse} etc. Multilayer networks not only have rich correlated structures~\cite{PRE,Vito_corr,Goh,Raissa} that encode more information than a single layer, but also contain various dynamical processes that are strongly affected by the multiplexity of the network. These dynamical processes include percolation~\cite{Buldyrev2010,Goh,Baxter2012,Cellai1,Algorithm,BD1,Cellai2}, diffusion \cite{Arenas1,Arenas2}, epidemic spreading \cite{Boguna_epidemics,Cozzo_epidemics,Arenas_aware}, and game theory
~\cite{game1,game2} etc.

Multiplex networks are a special class of multilayer network consisting of a set of nodes connected by $M$ different types of links. Each network consisting of a given type of link interaction forms one of the $M$ layers of a multiplex network.

Most social networks are multiplex. In fact, social ties have different connotations possibly indicating friends, colleagues, acquaintances and family relations, etc. Moreover, in the modern society, online social interactions can occur between different online social networks such as Twitter, Facebook and LinkedIn etc. The vast majority of data on multiplex social networks display a significant link overlap~\cite{Weighted, PRE, Thurner}. This property indicates that a significant fraction of pair of nodes can be connected at the same time by more than one type of interaction. For example, it might occur that a colleague is also a friend or that a two individuals might be connected at the same time in Facebook and Twitter.

The opinion dynamics on social multiplex networks have been investigated recently using spin models such as the voter models~\cite{Masuda,Marina1,Marina_Vito,Chmiel1,Chmiel2}, election models~\cite{Election}, and Hamiltonian spin systems~\cite{Vito_spin}. The observed dynamics on social multiplex networks cannot be reduced to the dynamics on a single social aggregated network that treats all the interactions of the multiplex network on an equal footing. In adaptive voter models, an absorbing and shattered fragmentation transition~\cite{Marina1,Marina_Vito} occurs in which one layer can be fragmented into two clusters each one reaching consensus on a different opinion, whereas the other layer remains connected in one cluster. In election models, the competing campaigns of two parties can give rise to election outcomes in which both parties have a large electorate \cite{Election}. Additionally, the party investing more in building a connected network of supporters is more likely to win the election \cite{Election}. In studies of the opinion dynamics on multiplex networks, where different opinion can be spread across different layers, an important question is whether each node maintains coherent behavior, that is, has a similar opinion in all the layers. A spin opinion model displaying a coherence--incoherence transition was numerically investigated recently~\cite{Vito_spin}. Spins are coupled within each layer to represent the interaction between one node and its neighbors on a given topic, and also across layers to represent the tendency of each node to take a coherent opinion on all the topics.

Another spin model, which illustrates opinion dynamics  in social networks due to the influence of interdependence between different social communities, is the Ashkin-Teller model~\cite{Ashkin1943}. It was studied on scale-free (SF) network in which the degree distribution follows power law and an analytical approach revealed that a rich phase diagram including the critical end point was obtained~\cite{AT}. It was considered on a duplex network with identical topology; however, the most realistic multiplex network~\cite{Buldyrev2010} can be the case in which the layers of the bilayer network are distinct and the amount of overlap is tunable.

Here our goal is to investigate to what extent link overlap affects the opinion dynamics defined on multiplex networks and whether link overlap favors coherent opinions. 

We consider a duplex network formed by two layers where a two-state opinion dynamics takes place. For example, one could consider a voting model for the city council and for the national parliament. For each vote, nodes can be influenced by a different set of nodes. In the previous example, the first layer indicates the network influencing the city council vote, the second layer indicates the network determining the national vote. The link overlap has a clear effect on this opinion dynamics by coupling the two layers. In fact if two nodes are connected in both layers it is natural to assume that the simultaneous alignment of the opinions in both layers must be favored by the dynamics. This considerations allow us to model the opinion dynamics in presence of link overlap, with a spin Hamiltonian model that is a variation of the Ashkin-Teller (AT) model~\cite{Ashkin1943,AT} that we call $g$-AT model. 

The model contains two species of Ising spins, the $s$-spin and $\sigma$-spin, with each species of spin located on a single layer of the duplex network. The duplex network is a maximum entropy duplex network with given multidegree distribution \cite{PRE} and as such it is very suitable to modulate the role of overlapping multilinks. In particular we here assume that non-overlapping multilinks and overlapping multilinks have a SF multidegree distribution characterized by a different power-law exponent. Here we provide a complete analytical mean-field solution to this model and we reveal the complex phase diagram of the model. We show that favoring the simultaneous alignment of the opinions of nodes connected in both layers provides a simple mechanism to generate coherence of opinions.

This paper is organized as follows: We introduce the Hamiltonian of the $g$-AT model and the duplex network topology under study in Sec II. In Sec. III we derive the free energy density using the mean-field approximation, and then self-consistency equations for the order parameters by minimizing the free energy density. Next, from these self-consistency equations, we  obtain the susceptibilities. In Secs. IV and V, we obtain rich phase diagrams in which different phases in the parameter space are delimited by lines indicating phase transitions {(PTs)} of different order. Note that the phase diagrams are richer than those of the original AT model on SF networks~\cite{AT}, because the links are classified into two types: non-overlapping and overlapping links. Finally, we summarize the results in Sec. VI.

\begin{figure}
\resizebox{1.0\columnwidth}{!}{\includegraphics{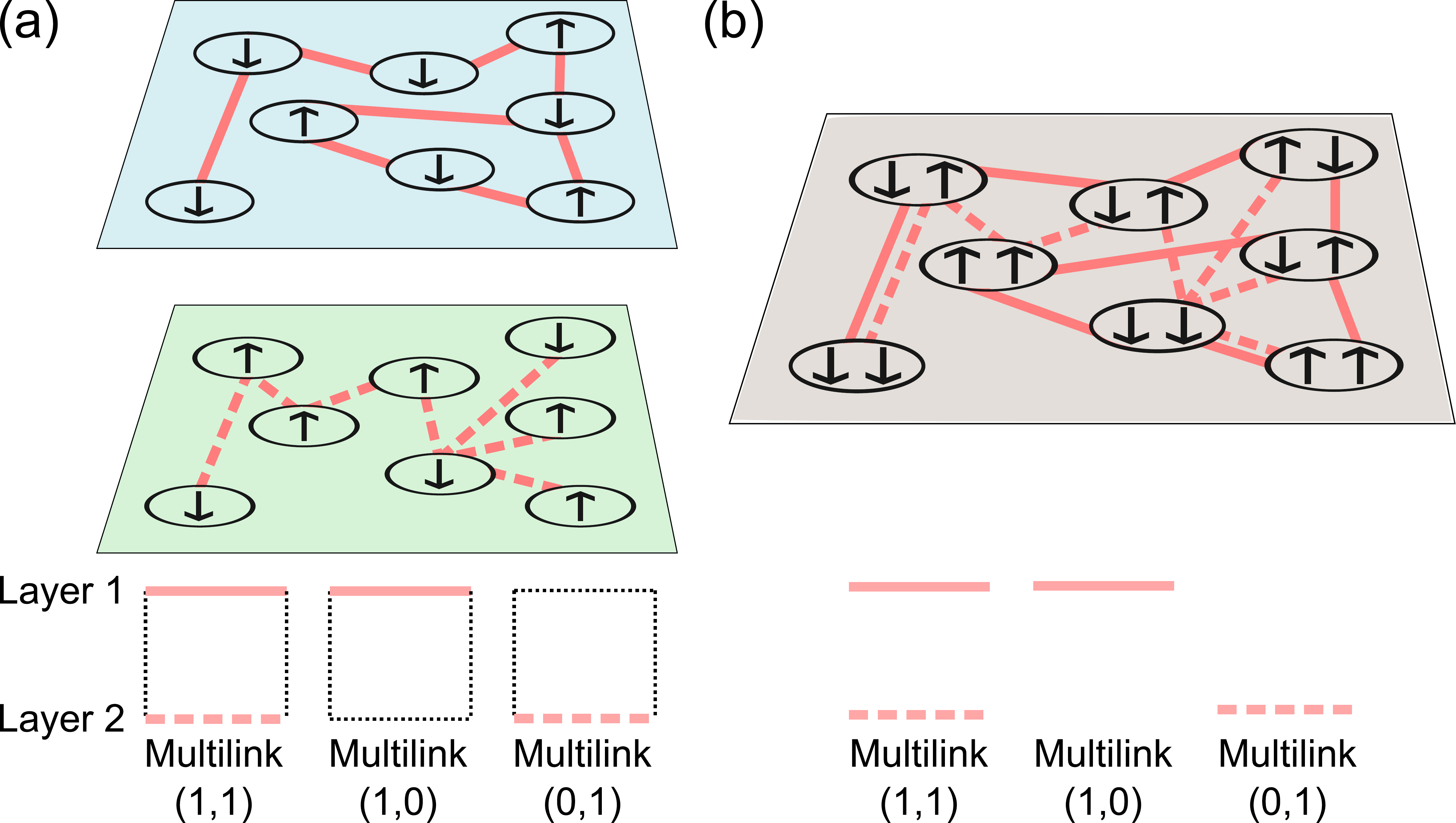}}
\caption{(Panel (a)) The $g$-AT model on a duplex network: two species ($s_i$, $\sigma_i$) of Ising spins describe respectively the opinion  of node $i$  in layer 1 and in layer 2.  Each pair of nodes of the duplex network can be connected by a different type of multilink: multilinks  $(1,1)$ connect pair of nodes in both layer 1 and layer 2; multilinks $(1,0)$ and $(0,1)$ connect pair of nodes only in layer 1 and only in layer 2 respectively.
Therefore multilinks $(1,1)$ describe overlapping links while multilinks $(1,0)$ and $(0,1)$ describe non overlapping links.
The model can be also interpreted as a model on a colored network in which nodes are associated pair of spin and the interactions between each pair of nodes can be distinguished in mutlilinks $(1,1),(1,0)$ and $(0,1)$ (panel (b)).
\label{fig:1a}	
}
\end{figure}

\begin{figure}
\resizebox{1.0\columnwidth}{!}{\includegraphics{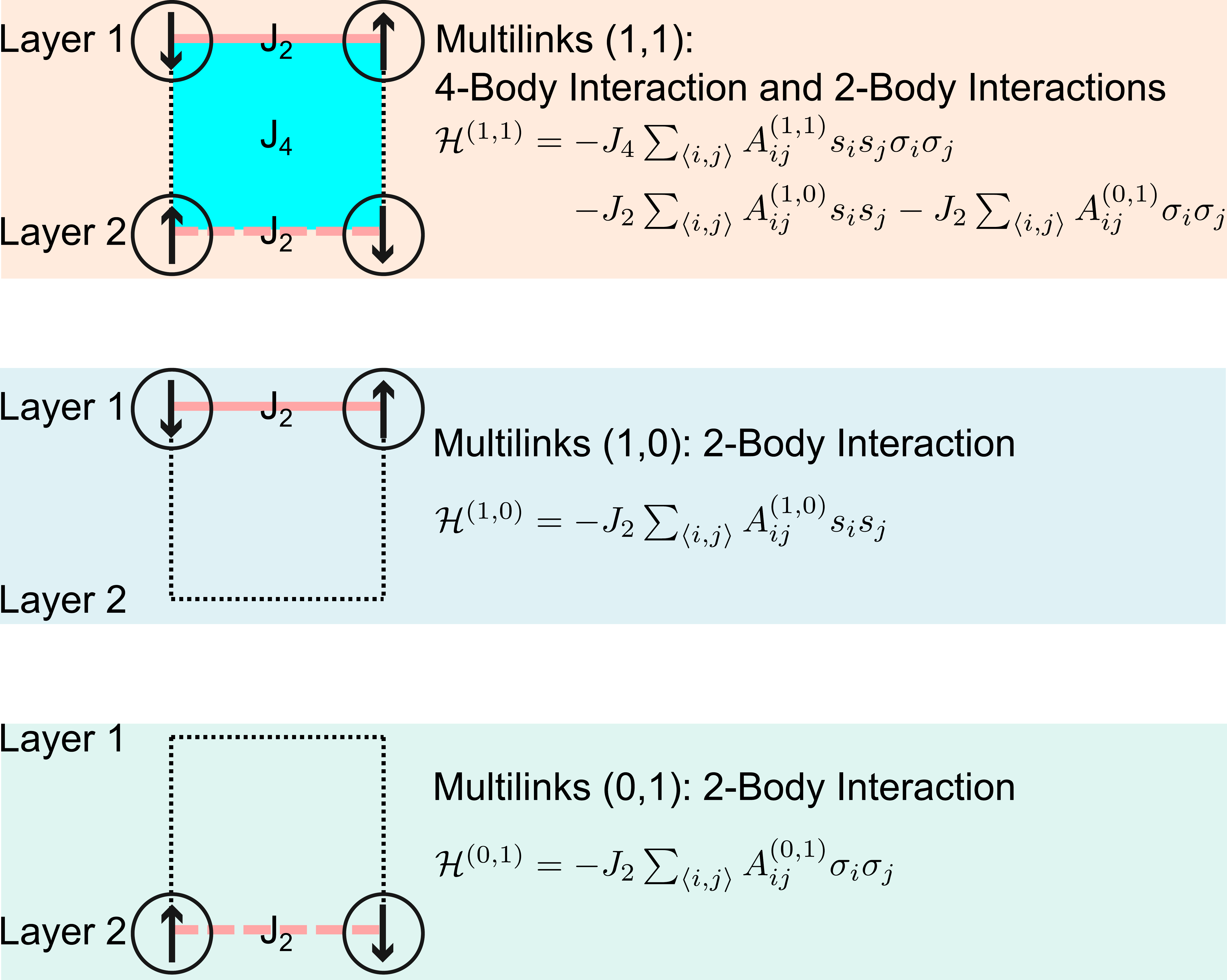}}
\caption{The $g$-AT model is an Hamiltonian model combining two-body and four-body interactions. The four body interactions characterizes the interactions between the spins of species  $s_i$, and of species  $\sigma_i$ connected by a multilink $(1,1)$. The two-body interactions characterize the coupling between spins of a given species (either the spins $s_i$ or the spins $\sigma_i$) connected by either a multilink $(1,0)$ or $(0,1)$.
\label{fig:1b}	
}
\end{figure}

\section{MODEL AND FORMALISM}\label{sec:main_result}

We consider a duplex network formed by $N$ nodes $i\in \{1,2,\ldots, N\}$. Every pair of nodes $(i,j)$ of the duplex network can be connected in multiplex ways. In order to indicate these different type of connections we use multilinks introduced in Ref.~\cite{PRE}.
In particular we say that a pair of nodes $(i,j)$ is connected by a multilink $(1,0)$ if they are only connected in layer 1, they are connected by a multilink $(0,1)$ if they are only connected in layer 2, and they are connected by a multilink $(1,1)$ if they are connected in both layers.
Every pair of nodes can be connected only by one type of multilink,  alternatively they can be unconnected in both layers (see Figure \ref{fig:1a}). 
We call multiadjacency matrices the matrices of  elements  $A^{(1,0)}_{ij}$, $A^{(0,1)}_{ij}$, and  $A^{(1,1)}_{ij}$ indicating whether or not the pair of nodes  $(i,j)$ is connected by a multilink $(1,0)$, a multilink $(0,1)$, and a multilink $(1,1)$, respectively. 
This general duplex network topology includes link overlap captured by the multilinks $(1,1)$. The presence of such multilinks has been observed in a variety of social networks \cite{Weighted,Thurner}.
Here and in the following we indicate with multidegrees $k_i^{(1,0)},k_i^{(0,1)}$, and $k_i^{(1,1)}$ the numbers of multilinks incident to the node $i$, i.e.,
\bea
k_i^{(1,0)}&=&\sum_{j=1}^NA_{ij}^{(1,0)},\nonumber \\
k_i^{(1,0)}&=&\sum_{j=1}^NA_{ij}^{(1,0)},\nonumber \\
k_i^{(1,1)}&=&\sum_{j=1}^NA_{ij}^{(1,1)}.
\eea

On such a duplex network, we consider the $g$-AT model that describes opinion dynamics and takes into account the role that link overlap has on this dynamics. We consider two species of Ising spins $s_i$ and $\sigma_i$ associated with the dynamics on layer 1 and layer 2, respectively. The two spins take values $s_i\in \{-1,1\}$ and $\sigma\in \{-1,1\}$. 
These spin variables are interacting via 2-body interactions and 4-body interactions (see Figure $\ref{fig:1b}$). In particular, for each multilink $(1,0)$ connecting node $i$ to node $j$, we have a 2-body Ising interaction between the spins $s_i$ and $s_j$ with coupling constant $J_2$. Similarly, for each multilink $(0,1)$ connecting node $i$ to node $j$, we have a 2-body Ising interaction between the spins $\sigma_i$ and $\sigma_j$ with coupling constant $J_2$. For each multilink $(1,1)$ instead we consider a combination of 2-body and 4-body interactions. The 2-body interactions tend to align spins associated to the same layer with coupling constant $J_2$. The 4-body interactions couples instead the four spins $s_i$, $s_j$, $\sigma_i$, and $\sigma_j$, and is modulated by a coupling constant $J_4$. In particular, the Hamiltonian of the $g$-AT model without an external magnetic field is expressed as the sum of three terms{,}
\bea
\mathcal{H}_o=\mathcal{H}^{(1,0)}+\mathcal{H}^{(0,1)}+\mathcal{H}^{(1,1)}
\eea
where
\bea
\mathcal{H}^{(1,0)}&=&-J_2\sum_{\langle i,j\rangle}A_{ij}^{(1,0)}s_is_j\nonumber \\
\mathcal{H}^{(0,1)}&=&-J_2\sum_{\langle i,j\rangle}A_{ij}^{(0,1)}\sigma_i\sigma_j\nonumber \\
\mathcal{H}^{(1,1)}&=&-J_4\sum_{\langle i,j\rangle }A_{ij}^{(1,1)}s_is_j\sigma_i\sigma_j\nonumber\\ 
&&-J_2\sum_{\langle i,j\rangle}A_{ij}^{(1,0)}s_is_j-J_2\sum_{\langle i,j\rangle}A_{ij}^{(0,1)}\sigma_i\sigma_j
\eea
with the pairs of connected nodes $\langle i,j\rangle$. 
Alternatively, the Hamiltonian $\mathcal{H}_o$ of the $g$-AT model without an external magnetic field can be expressed more concisely as
\begin{align}
-\beta\mathcal{H}_o
				= & K_2\sum_{\langle i,j \rangle} \mathbf{s}_i^\Tr \textbf{A}_{ij} \mathbf{s}_j \,,
				\label{eq:hamiltonian_beta}
\end{align}
where $\mathbf{s}_i=(s_i,\sigma_i,s_i\sigma_i)^\Tr$, and the matrix $\textbf{A}_{ij}$ is given by
\begin{equation}
\label{eq:matrix}
{\bf A}_{ij}=\left(\begin{array}{ccc} 
A^{(1,0)}_{ij}+A^{(1,1)}_{ij} & 0 & 0 \\ 
0 & A^{(0,1)}_{ij}+A^{(1,1)}_{ij} & 0 \\ 
0 & 0 &  xA^{(1,1)}_{ij}
\end{array}\right),
\end{equation}
where $x\equiv J_4/J_2$. 
Moreover, $\beta=1/k_{\rm B}T$, where $k_{\rm B}$ is the Boltzmann constant, $T$ is the temperature, and $K_2 \equiv \beta J_2$ with coupling constant $J_2$. For later discussion, we define similarly $K_4\equiv \beta J_4$. 

Here, we investigate the critical properties of this model on a maximum entropy duplex network model with given multidegree distribution \cite{PRE}. In order to distinguish between multilinks $(1,1)$ which imply link overlap and the other multilinks $(1,0)$ and $(0,1)$ which do not,  we assume for simplicity that  each node $i$ of the multiplex network has the same multidegree $(1,0)$ and multidegree $(0,1)$, and we indicate the multidegree of non-overlapping multilinks and of overlapping multilinks as 
\bea
k_i^{(1,0)}&=&k_i^{(0,1)}=k_{\n,i} \,,\nonumber \\
k_i^{(1,1)}&=&k_{\o,i}.
\eea
where the subscript $\n$ of $k_{\n,i}$ indicates and the subscript $\o$ of $k_{\o,i}$ indicates {\it non-overlap} and {\it overlapping} multilinks, respectively. We assume that the degree distributions corresponding to {\it overlapping} and {\it non-overlapping} multilinks are power-law functions with exponents $\lambda_\o$ and $\lambda_\n$, respectively. The degree distribution is shortly written as 
\bea 
P_d(k_a)\sim k_a^{-\lambda_a},
\eea 
where $a\in \{\o,\n\}$.

In the considered ensemble of duplex networks \cite{PRE} a pair of nodes $(i,j)$ is connected by $(1,0)$ multilinks with probability $p^{(1,0)}_{ij}$, by $(0,1)$ multilinks with probability $p^{(0,1)}_{ij}$, and by $(1,1)$ multilinks with probability $p^{(1,1)}_{ij}$, where we have 
\bea
p^{(1,0)}_{ij}&=&\frac{k_i^{(1,0)}k_j^{(1,0)}}{\Avg{k^{(1,0)}}N},\nonumber \\
p^{(0,1)}_{ij}&=&\frac{k_i^{(0,1)}k_j^{(0,1)}}{\Avg{k^{(0,1)}}N},\nonumber \\
p^{(1,1)}_{ij}&=&\frac{k_i^{(1,1)}k_j^{(1,1)}}{\Avg{k^{(1,1)}}N}
\label{eq:p_marginals}
\eea
with $\langle k^{(1,0)} \rangle$, $\langle k^{(0,1)} \rangle$, and $\langle k^{(1,1)} \rangle$ being the average multidegrees.
Indeed these  marginal probabilities are obtained in the maximum entropy ensemble with given multidegree distribution as long as the degree distribution display the structural cutoff. Here, we consider the thermodynamic limit ($N \to \infty$) and power-law exponents greater than $3$, so that the effect of structural cutoff can be ignored. 

The phase diagram of this model will be affected by the topology of multiplex network and the strength of the interlayer interaction. This can be studied as a function of three parameters, $\lambda_\n$, $\lambda_\o$, and $x\equiv J_4/J_2$. The ratio $x$ quantifies the degree strength of 4-body interaction with respect to the strength of 2-body interactions for between nodes linked by multilinks $(1,1)$.

The original AT model ~\cite{AT} comprises  two species of Ising spins, $s_i$ and $\sigma_i$, locating at each node $i$ on a monolayer network. 
The original AT-model can be thus recovered as a limit case of the   $g$-AT model  in absence of non-overlapping multilinks and when $x=1$ (i.e. $J_4=J_2$). 
{Indeed} in this limit we recover the  Hamiltonian for the original AT model given by 
\begin{align}
\mathcal{H}=-J_2\sum_{\langle ij \rangle}(s_is_j+\sigma_i\sigma_j)-J_2\sum_{\langle ij\rangle} s_is_j\sigma_i\sigma_j,
\end{align}
that can be rewritten in the form of the 4-state Potts model as 
\begin{align}
\mathcal{H}=-4J_2\sum_{\langle ij \rangle}\left(\delta(q_i, q_j)-1/4\right),
\end{align}
where $q_i$ is a Potts {spin with value} $0,1,2$ or $3$ at node $i$ and $\delta(q_i,q_j)=1$ for $q_i=q_j$, and zero otherwise~\cite{Kadanoff}.
Since the non-overlapping multilinks are absent, the phase diagram of the original model is a function of a single power-law  exponent $\lambda$ of the degree distribution. Clearly this power-law exponent correspond to the power-law exponent $\lambda_\o$ of overlapping links of the $g$-AT model.

\section{Mean-field solution}
To obtain the Landau free energy, we calculate the Hamiltonian in Eq.~\eqref{eq:hamiltonian_beta} by the mean-field approximation. We first take the local order parameters $\mathbf{m}_i=(m^s_i, m^{\sigma}_i, m^{s\sigma}_i)^\Tr$, whose components are defined as $m^s_i =\langle s_i \rangle$, $m^\sigma_i =\langle \sigma_i \rangle$, and $m^{s\sigma}_i =\langle s_i\sigma_i \rangle$. 
Here $\langle \cdots \rangle$ is the ensemble average of a given quantity. Next, we expand each spin variable with respect to the respective local order parameter as $\mathbf{s}_i=(m^s_i + \delta m^s_i,m^\sigma_i + \delta m^\sigma_i,m^{s\sigma}_i + \delta m^{s\sigma}_i)^\Tr$. 
We can neglect the higher-order terms in $\delta m^s_i$, $\delta m^\sigma_i$, and $\delta m^{s\sigma}_i$ because the magnitude of these terms is very small compared to that of the local order parameter. The mean-field Hamiltonian $\mathcal{H}_{\textrm{mf}}$ can be written as 
\begin{align}
-\beta\mathcal{H}_{\textrm{mf}} \simeq
&- K_2\sum_{i,j} \mathbf{m}_i^\Tr \textbf{A}_{ij} \mathbf{m}_j + K_2\sum_{i,j} \mathbf{m}_i^\Tr \textbf{A}_{ij} (\mathbf{s}_j + \mathbf{\sigma}_j) \,.
\label{eq:hamiltonian_beta_meanfield}
\end{align}
Then, we obtain the mean-field Landau free energy $\mathcal{F}$, which is given by
\begin{align}
	\beta\mathcal{F} 
	&= -\ln Z \cr
	& =-\ln\sum_{\{s_i,\sigma_i\}}e^{-\beta\mathcal{H}_{\rm mf}} \simeq -\sum_{i}\ln Z_i + K_2\sum_{i,j} \mathbf{m}_i^\Tr \textbf{A}_{ij} \mathbf{m}_j,
	\label{eq:freeEnergy}
	\end{align}
where	
	\begin{align}
	Z_i 
	= 4\left[\C_i(s)\C_i(\sigma)\C_i(s\sigma) +\S_i(s)\S_i(\sigma)\S_i(s\sigma)\right],
	\label{eq:freeEnergy_Z}
		\end{align}
	with 
	\begin{align}
	\C_i(s)\equiv \cosh\left(\sum_{j\in \text{nn}(i)}K_2 m^s_j\right),\quad \S_i(s)\equiv \sinh\left(\sum_{j\in \text{nn}(i)}K_2 m^s_j\right)\,.
	\end{align}
Here $\sum_{j\in \text{nn}(i)}$ indicates that the summation runs over all the nearest neighbors $j$ of node $i$ for each of the three types of links.	

Next, we use the annealed approximation to perform the summation: 
\begin{align} 
\left.
\begin{array}{ll}
\sum_{\langle i,j\rangle} A^{(1,0)}_{ij}\mathcal{A}_{ij} \rightarrow \frac{1}{2}\sum_{i, j} p_{ij}^{(1,0)} \mathcal{A}_{ij}\,, \\ \\
\sum_{\langle i,j\rangle} A^{(0,1)}_{ij}\mathcal{A}_{ij} \rightarrow \frac{1}{2}\sum_{i, j}  p_{ij}^{(0,1)}\mathcal{A}_{ij}\, \textrm{ and} \\ \\
\sum_{\langle i,j\rangle} A^{(1,1)}_{ij}\mathcal{A}_{ij} \rightarrow \frac{1}{2}\sum_{i, j} p_{ij}^{(1,1)} \mathcal{A}_{ij}\,,
\end{array} \right.
\end{align}
where  $\mathcal{A}_{ij}$ is a given function of $i$ and $j$ and $p_{ij}^{(1,0)}, p_{ij}^{(0,1)}$ and $p_{ij}^{(1,1)}$ are defined in Eq. (\ref{eq:p_marginals}). 

We define a global order magnetization for $s$ spin : 
\begin{align} 
m^{(1,0)}_s  =  \dfrac{\sum_{i}k_i^{(1,0)} m^s_i}{N\langle k^{(1,0)} \rangle} \textrm{ and } m^{(1,1)}_s  =  \dfrac{\sum_{i}k_i^{(1,1)} m^s_i}{N\langle k^{(1,1)} \rangle},
\end{align}
where $m^s_i$ is the local order parameter for $s$ spin. We introduce global order parameters for $\sigma$ and $s\sigma$ spins similarly. Then, we set that $M\equiv m^{(1,1)}_{s\sigma}${.}

Since the considered duplex network ensemble has the same multidegree distribution of the non-overlapping multilinks we can set 
	\begin{align}
	m_s^{(1,0)} = m_\sigma^{(0,1)} \equiv m_\n \,,\qquad
	\textrm{ } m_s^{(1,1)} = m_\sigma^{(1,1)} \equiv m_\o \,.
	\end{align}
The three order parameters are now denoted as $m_\o $, $m_\n $, and $M$-magnetization, respectively. 
Applying the annealed approximation, we rewrite the free energy density ($f \equiv \beta \mathcal{F} / N$) in terms of the order parameters $m_\n $, $m_\o $, and $M$. The free energy density $f$ is given by
	\begin{align}
	&f \simeq K_2 m_\n^2 \langle k_\n\rangle + K_2 m_\o^2 \langle k_\o\rangle + \dfrac{1}{2}K_4 M^2 \langle k_\o \rangle \cr
	&~ - 2\int_{k_{\rm min}^\n}^{\infty}\int_{k_{\rm min}^\o}^{\infty} dk_\n dk_\o P_d(k_\n) P_d(k_\o) \ln\left[\cosh\left(K_2 (m_\n k_\n + m_\o k_\o)\right) \right] \cr
	&~ - \int_{k_{\rm min}^\o}^{\infty} dk_\o P_d(k_\o) \ln\left[\cosh\left(K_4 M k_\o\right) \right] -\B_1 \,,
	\label{eq:freeEnergyDensity}
	\end{align}
where $K_4=\beta J_4$ with coupling constant $J_4$ and  
	\begin{align}
	\B_1 
	&= \int_{k_{\rm min}^\n}^{\infty} \int_{k_{\rm min}^\o}^{\infty} dk_\n dk_\o P_d(k_\n) P_d(k_\o) \ln\left(1+\T_2^2\T_4 \right) \,,
	\label{eq:B}
	\end{align}
with 
	\begin{align}
	\T_2 \equiv \tanh\left(K_2 (m_\n k_\n + m_\o k_\o)\right), \, \T_4\equiv \tanh\left(K_4 M k_\o\right)\,.
	\end{align}

Minimizing the free energy density $f$, ${\partial f}/{\partial m_ a  }=0$ and ${\partial f}/{\partial M}=0$, we obtain the following self-consistency relations:
	\begin{align}
	m_ a  \langle k_ a   \rangle 
	&= \int_{k_{\rm min}^\n}^{\infty} \int_{k_{\rm min}^\o}^{\infty} dk_\n dk_\o P_d(k_\n) P_d(k_\o) \dfrac{\T_2\left(1+\T_4 \right)}{1+\T_2^2\T_4} k_ a   
	\label{eq:equationOfState_m}
	\end{align}
where $a \in \{ \o, \n\}$, and
	\begin{align}
	M\langle k_\o \rangle
	&= \int_{k_{\rm min}^\n}^{\infty} \int_{k_{\rm min}^\o}^{\infty} dk_\n dk_\o P_d(k_\n) P_d(k_\o) \dfrac{\T_4+\T_2^2}{1+\T_2^2\T_4} k_\o .
	\label{eq:equationOfState_M}
	\end{align}
The self-consistency relations~\eqref{eq:equationOfState_m} and~\eqref{eq:equationOfState_M} admit three solutions,  corresponding to the paramagnetic phase ($m_a=0,M=0$), the Baxter phase ($m_a>0, M>0$), and the $\langle \sigma s\rangle$ phase ($m_a=0, M>0$). 

To obtain the susceptibility, we also consider a Hamiltonian including an external magnetic field, given by 
\begin{align}
-\beta\mathcal{H}
	&= - \beta\mathcal{H}_o + \sum_{i} \mathbf{H}_{i}^{\Tr} \mathbf{s}_i\,,
\label{eq:hamiltonian_ext_field}
\end{align}
where $\textbf{H}_{i}=(k_{\o,i} H_\o +k_{\n,i} H_\n , k_{\o,i} H_\o+k_{\n,i} H_\n , k_{\o,i} H_4)^\Tr$. $H_a$ is the external magnetic field applied to $s$ and $\sigma$ spins in proportion to the multidegree $k_{\a}$ and $H_4$ is another external magnetic field applied to $s \sigma$ spins in proportion to degree $k_\o$. Minimizing the free energy density, we obtain the self-consistency equations for magnetizations with respect to external magnetic fields:
\begin{equation}
- \dfrac{\partial f}{\partial H_a} = m_a \langle k_a \rangle, \quad - \dfrac{\partial f}{\partial H_4} = M \langle k_\o \rangle\,. \label{eq:equationOfState_ext_field}
\end{equation}
These self-consistency equations can be obtained by substituting $K_2 m_a k_a$ and $K_4 M k_\o $ with $(K_2 m_a +H_a)k_a$ and $(K_4 M +H_4)k_\o $, respectively, in Eqs.~\eqref{eq:freeEnergyDensity}, \eqref{eq:B}, \eqref{eq:equationOfState_m}, and \eqref{eq:equationOfState_M} (see Appendix A). 
The susceptibilities are calculated using the following relations:
\begin{equation}
\chi_a \equiv \dfrac{\partial m_a}{\partial H_a}\Big|_{H_a,H_4 \to 0}, \quad \chi_M \equiv \dfrac{\partial M}{\partial H_4}\Big|_{H_a,H_4 \to 0}\,.
\end{equation}
Using the above relations, the susceptibilities $\chi$ can be obtained as follows:
\begin{align}
\chi_a   
& = \frac{\mathcal{A}_{aa} + \mathcal{A}_{a \bar a} K_2 \partial m_{\bar a}/\partial H_a   +  \mathcal{A}_{a M} K_4 \partial M/\partial H_a}{\langle k_a \rangle - K_2 \mathcal{A}_{aa}} \,, \\
\chi_M 
& = \frac{\mathcal{A}_{MM} + \mathcal{A}_{M \o} K_2 \partial m_\o /\partial H_4 +  \mathcal{A}_{M \n} K_2 \partial m_\n /\partial H_4}{\langle k_\o \rangle - K_4 \mathcal{A}_{MM}} \,,
\label{chi_explicit}
\end{align}
where $a\in \{\n,\o\}$ and  $\bar a \in \{\n,\o\}$ with $a$ different from $\bar{a}$. Here the  $\mathcal{A}$ terms are obtained as follows
\begin{align} 
\mathcal{A}_{aa} & = \dfrac{\partial m_{a\I}}{\partial H_a}, \quad \, \mathcal{A}_{a\bar a} = \dfrac{\partial m_{a\I}}{\partial H_{\bar a}}, \quad \,
\mathcal{A}_{aM} =  \dfrac{\partial m_{a\I}}{\partial H_4}, \, \cr
\mathcal{A}_{Ma} & = \dfrac{\partial M_\I}{\partial H_a}, \quad \hbox{and} \quad \mathcal{A}_{MM} = \dfrac{\partial M_\I}{\partial H_4}.
\end{align}
In  Appendix B,  we provide the extensive formulas for $\mathcal{A}$s  in the limit $H_a\to 0$ and $H_4 \to 0$, where $m_{a\I}$ and $M_\I$ are presented in intergral form in Eqs.~\eqref{eq:equationOfState_m_H} and Eqs.~\eqref{eq:equationOfState_M_H}.

\section{Phase diagram I: $J_4/J_2$-DEPENDENCE}\label{method}

\subsection{Phases of the model}

The $g$-AT model has one of the three phases, paramagnetic phase, Baxter phase, and $\langle \sigma s\rangle$ phase, in equlibrium state, depending on $\lambda_\n$, $\lambda_\o$, $x=J_4/J_2$, and $T$.
\begin{itemize} 
\item[--] The paramagnetic phase is characterized by the order parameters $m_a=\avg{s}=\Avg{\sigma}=0, M=\avg{\sigma s}=0$. This is the characteristic phase found in the  high temperature region, where the stochastic element of the dynamics is dominant. This phase corresponds to an equilibrium configuration in which there is no majority opinion in either layer ($m_a=0$), and each node has a random and uncorrelated opinion in the two different layers ($M=0$). Therefore this is the phase entirely dominated by noise.

\item[--]The Baxter phase is characterized by the order parameters $m_a=\avg{s}=\avg{\sigma}>0$, and $M=\avg{\sigma s}>0$.
This is the phase in which we observe the formation of a majority opinion which is the same in both layers ($m_a>0$). Therefore, each node has coherent opinions  in the two distinct layers ($M>0$).

\item[--] The $\avg{\sigma s}$ phase (Coherent phase) is characterized by the order parameters $m_a=\avg{s}=\avg{\sigma}=0$, and $M=\avg{\sigma s}>0$. This phase occurs for high temperature and when $x=J_4/J_2$ is sufficiently high, in which the 4-body interactions are stronger than the 2-body interactions; therefore each single node of the multiplex network tends to have the same opinion in both layers but these opinions are not yet aligned with the opinion of their neighbours. As a consequence there is no yet formation of a majority opinion in each  layer (i.e. $m_a=0$). Note that the term of $\avg{\sigma s}$ phase originates from the original paper in physics~\cite{Ashkin1943}. To impose a meaning on the phase in the perspective of opinion formation, we call it Coherent phase hereafter.
\end{itemize}

\subsection{Classification of critical points and regions of the phase diagram}

In the $g$-AT model the transitions between the phases Para, Baxter and Coherent occurs as a function of the temperature $T$ and very diverse and rich critical phenomena are observed. Indeed the PTs can be continuous, discontinuous, hybrid and in general we can observe more than one PT as the temperature $T$ is lowered, while the other parameters are kept unchanged.

\begin{figure}
\resizebox{0.95\columnwidth}{!}{\includegraphics{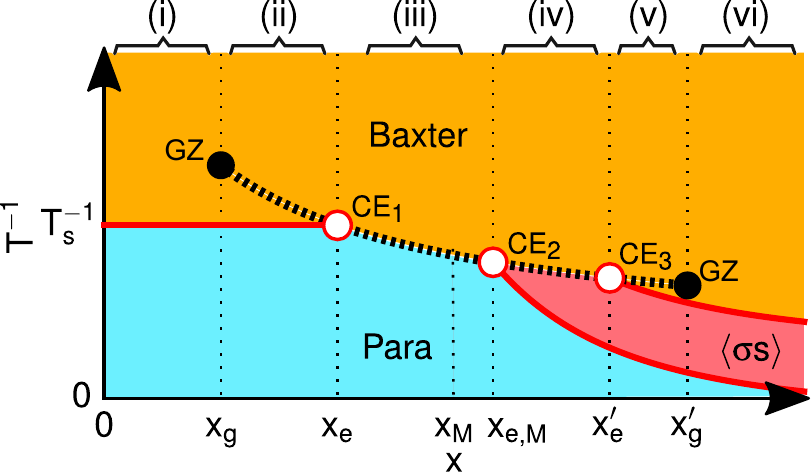}}
\caption{Schematic phase diagram of the $g$-AT model for a given set of $\lambda_\n = 3.53$ and $\lambda_\o = 3.90$. Solid and dotted curves represent continuous and discontinuous PTs, respectively. This phase diagram is mostly similar to the one of the original AT model~\cite{AT}.
\label{fig:fig3}
}
\end{figure}

To be concrete we discuss here an exemplar phase diagram of the  $g$-AT model  in the parameter space [$x, T^{-1}$] (see Fig.~\ref{fig:fig3}). This phase digram is  obtained for the power-law exponents $\lambda_\n =3.53$ and $\lambda_\o =3.90$. For this value of the power-law exponents, the phase diagram is similar to that of the original AT model~\cite{AT} in the range $\lambda_c < \lambda_\o < 4$, where $\lambda_c\approx 3.503$ indicates the  tricritical point (TP) of the  original AT model~\cite{AT}. In particular we recall that in the original AT-model for $\lambda_\o > \lambda_c$, the PT is of the first-order; otherwise, it is of the second-order~\cite{AT,Potts,Potts_mendes}. Here we will describe in detail this phase diagram while the dependence of the phase diagram on the power-law exponents $\lambda_\n$ and $\lambda_\o$ will be treated in the next section.

In the phase diagram shown in Fig.~\ref{fig:fig3}, the three phases of the dynamics are denoted by  Para ($m_a=0,M=0$),  Baxter phase ($m_a>0, M>0$), and  $\langle \sigma s \rangle$ phase ($m_a=0, M>0$). Dotted and solid lines represent discontinuous and continuous PTs, respectively. The critical temperature $T_s$ denotes the temperature at which a  second-order PT occurs from the Para phase to the Baxter phase.  Note that {$T_s$ is independent on} $x=J_4/J_2$ for $x < x_e$. 

The phase diagram has characteristic points denoted as GZs and CEs. We indicate with GZ a point at which the jump size (gap) of the order parameter becomes zero at {each side of the dotted} curve. We indicate with CE a critical endpoint, locating at the end of a continuous PT line, at which a line of first-order PT and a line of discontinuity of the order parameter merge. We will show that a mixed-order (or hybrid) transition occurs at these CE points. There are two GZs and three CEs in Fig.~\ref{fig:fig3}. Their $x$ positions are asymmetric.

In the phase diagram shown in Fig.~\ref{fig:fig3} we distinguish six regions based on the ratio $x=J_4/J_2$: 

\begin{itemize}
\item[--] In region (i), as the temperature is lowered the system undergoes a continuous PT  at $T_s$ from the Para to the Baxter phase. Therefore as the noise is reduced the system {goes continuously} from a phase with the absence of any order, to a state with a clear majority opinion which is the same in both layers.  This transition is denoted as (i)-type PT.  
\item[--] In region (ii), a continuous PT occurs at {$T_s$} between the Para and the Baxter phase. As {$T$} is lowered further, a discontinuous jump of the order parameters $m_a$ and $M$ occurs subsequently at $T_f$, in which we observe a discontinuity in $m_a$ and $M$ between two non-zero values. This indicates that at $T_f$, there is a discontinuous increment in the fraction of nodes adopting the majority opinion. This transition is denoted as (ii)-type PT.
\item[--] In region (iii), a discontinuous PT occurs at $T_f$ between the Para and the Baxter phase. This implies that a majority opinion is formed abruptly in both layers. This transition is denoted as (iii)-type PT. 
\item[--] In region (iv), a continuous PT occurs between the Para and the Coherent  phase at $T_{s,M}$. As $T$ is decreased further, a discontinuous PT occurs at $T_f$ from the Coherent phase to the Baxter phase. This implies that as the noise is reduced, at temperatures below the first continuous transition each single node tend to adopt a coherent opinion in both layers, and then when the temperature is further reduced a majority opinion is reached abruptly in both layers. These transitions are denoted as (iv)-type PT. 
\item[--] In region (v), two continuous PTs occur successively: between the Para and the Coherent phase at $T_{s,M}$ and between the Coherent and the Baxter phase at $T_s'$, respectively. Then as the temperature is decreased further, the order parameters $m_a$ and $M$ jumps at $T_f$ from one finite value to another. These transitions are denoted as (v)-type PT.
\item[--] In region (vi), two continuous PTs occur between the Para and the Coherent phase at $T_{s,M}$ and between the Coherent and the Baxter phase at $T_s'$. These transitions are denoted as (vi)-type PT.
\end{itemize}

\subsection{Free-energy landscape for the $x$-dependence of phase transitions}

In this paragraph we will discuss the critical behavior of the $g$-AT model as a function of the parameter $x=J_4/J_2$.

The phase diagram of the $g$-AT model can be treated separately for $x<x_M$ and $x>x_M$, where $x_M$ indicates the characteristic ratio between $J_4$ and $J_2$. For $x < x_M$, the $J_2$ interactions are dominant, and $O(m_a) \gg O(M)$ near the transition temperature. On the other hand, for $x > x_M$, $J_4$ interactions (interlayer interaction) become dominant and $O(m_a) \ll O(M)$ near the transition temperature. Thus, the Coherent phase can emerge. For the original AT model, $x_M=1$; however, for the $g$-AT model, $x_M$ depends on $\lambda_\n$ and $\lambda_\o$. $x_M$ locates between $x_e$ and $x_{e,M}$ in Fig.~\ref{fig:fig3}. Explicitly formula to derive $x_M$ will be presented in Eq.~\eqref{eq:x_M}.

\begin{figure*}
\resizebox{2.0\columnwidth}{!}{\includegraphics{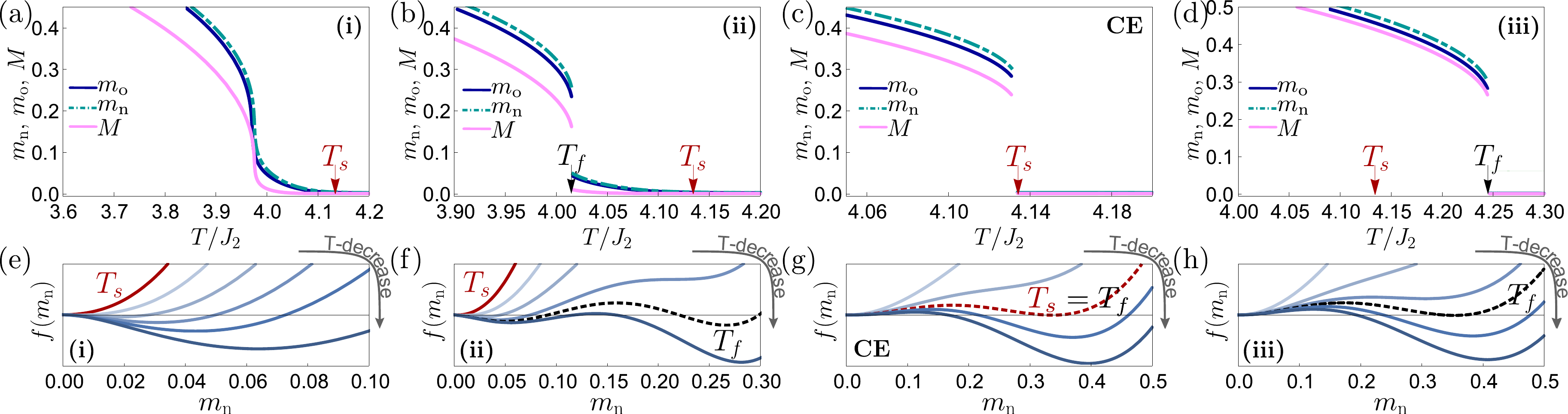}}
\caption{(a)--(d) Plot of the order parameters $m_a$ and $M$ as a function of $T/J_2$, and (e)--(h) Plot of the free energy density landscape as a function of $m_\n$ for $\lambda_\n=3.53$ and $\lambda_\o=3.90$ and various interlayer interaction ratios: $x=1.30$ for (a) and (e); $x=1.40$ for (b) and (f); $x=1.62$ for (c) and (g); and $x=1.80$ for (d) and (h). The transition types are second-order in region (i) (a) and (e); successive continuous-discontinuous in region (ii) for (b) and (f); mixed-order at CE$_{1}$ for (c) and (g); and discontinuous transition in region (iii) for (d) and (h).
\label{fig:fig4}	
} 	
\end{figure*}

\begin{figure*}
\resizebox{2.0\columnwidth}{!}{\includegraphics{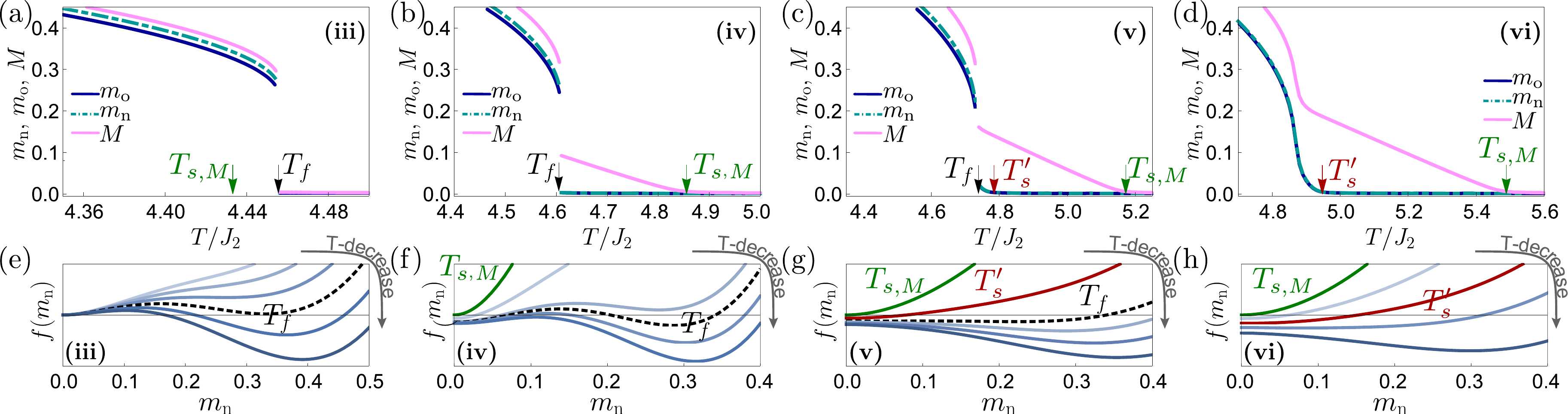}}
\caption{(a)--(d) Plot of the order parameters $m_a$ and $M$ as a function of $T/J_2$. (e)--(h) Plot of the free energy density landscape as a function of $m_\n$ for $\lambda_\n=3.53$ and $\lambda_\o=3.90$ and various interlayer interaction ratios: $x=2.10$ for (a) and (e); $x=2.30$ for (b) and (f); $x=2.45$ for (c) and (g); and $x=2.60$ for (d) and (h). The transition types are first-order in regime (iii) for (a) and (e); successive continuous-discontinuous in regime (iv) for (b) and (f); successive continuous-discontinuous in regime (v) for (c) and (g); and continuous transition in regime (vi) for (d) and (h).}
\label{fig:fig5}	
\end{figure*}

{The free-energy landscape determines the location and type of PTs with respect to $x$.} Here we provide the discussion of the main results obtained by investigating the properties of the free-energy density illustrated by in Figs.~\ref{fig:fig4} and \ref{fig:fig5} near $T_s$. We refer the interested reader to the exact formula of the free-energy density $f$ given in Appendix E.

\subsubsection{Case $0<x<x_M$}
In this paragraph we describe the critical behavior of the $g$-AT model for $x<x_M$ including regions (i), (ii) and (iii), and a point CE$_1$. In region (i) a continuous PT occurs between the Para and the Baxter phase. Therefore as the temperature is lowered, both $m_a$ and $M$ increase continuously for $T<T_s$.

Using Eqs.~\eqref{eq:free_energy_x<1_o} and \eqref{eq:free_energy_x<1_n}, we can obtain the critical behavior of the order parameters for $T<T_s$
\begin{align}
m_a &\sim \,  (T_s-T)^{\beta_m} && \quad \hbox{with} & \beta_m &= \dfrac{1}{\lambda_\min-3}, \\
M &\sim \, (T_s-T)^{\beta_M} && \quad \hbox{with} & \beta_M &= \dfrac{\lambda_\o -2}{\lambda_\min-3},
\label{eq:beta_exponent_x<1}
\end{align}
where $\lambda_{\min} = \min(\lambda_\o, \lambda_\n)$. 
The specific heat scales as  
\begin{align}
C \, \sim (T_s-T)^{-\alpha} && \quad \hbox{with} && \alpha = \dfrac{\lambda_\min-5}{\lambda_\min-3}.
\label{eq:alpha_exponent_x<1}
\end{align}
The susceptibility diverges as (see Appendix E for the derivation)
\begin{align}
\chi_a \sim 
\left\{\begin{array}{cccc}
(T_s-T)^{-\gamma^-} & \quad \hbox{with} \quad \gamma^-= 1 & \hbox{for} & T < T_s, \cr \cr
(T-T_s)^{-\gamma^+} & \quad \hbox{with} \quad \gamma^+= 1 & \hbox{for} & T > T_s .
\end{array} \right. 
\label{eq:susceptibility_m_<}
\end{align}

Secondly we observe that as $x$ is increased but still remains less than $x_M$, a jump arises in the order parameters $m_a$ and $M$ in region (ii), observed for  $x_g < x < x_e$ in Fig.~\ref{fig:fig3}. We observe that the system undergoes a continuous second order transition at $T_s$ between the Para and the Baxter phase characterized by the same critical exponents listed above. Moreover as the temperature $T$ is lowered further the system undergoes a sudden increase of the order parameter at $T_f$. Indeed at  $T_f$ the free energy density $f$ displays a global minimum at a finite $m_a$, leading to the abrupt change of the order parameters (see  Fig.~\ref{fig:fig4}(b) and (f)).

For  $x \to x_e$, the temperature $T_s$ becomes equal to  $T_f$. Therefore, two global minima of $f(m_a)$ occur at $m_a=0$ and $m_a >0$, simultaneously. At this point, the second-order and the first-order transition lines merge. Therefore, the critical behavior appears, together with the jump of the order parameters $m_a$ and $M$ as illustrated in Fig.~\ref{fig:fig4} (c) and (g). This type of PT is referred to as a mixed-order (or hybird) transition and this point is named as critical endpoint.  However, the susceptibility $\chi_m$ diverges at $T_s^+$ as it appears in a continuous PT.

In region (iii), for  $x_e < x < x_M$, a single  discontinuous transition occurs at $T_f$ between the Para and the Baxter phase. The critical bahavior in this region is illustrated   in Fig.~\ref{fig:fig4}(d) and (h). 

\subsubsection{Case $x \ge x_M$ }

Here, we consider the critical behavior of the $g$-AT model for $x \ge x_M$ including the regions (iii), (iv), (v) and (vi) and two critical endpoints (CE$_2$, CE$_3$).

At $x=x_M$ in region (iii), the free-energy densities $f(m_a)$ and $f(M)$ develop a local minimum at $m_a^*$ larger than 0 and at a temperature $T_f$ larger than $T_{s,M}$ (Fig.~\ref{fig:fig5}(a) and (e)). A discontinuous transition of $m_a$ and $M$ occurs between the Para and the Baxter phase at $T_f$. This phenomenology remains unchanged for $x_M<x<x_{e,M}$. 

At $x=x_{e,M}$, the system is at the  boundary between the regions (iii) and (iv), and we find the critical endpoint denoted as CE$_2$ in Fig.~\ref{fig:fig3}. At CE$_2$, when the temperature $T$ approaches $T_{s,M}$ from below, i.e., when we explore the critical behavior for $T\to T_{s,M}^-$, we observe a discontinuity in the value of $M$ from a non-zero value to zero, and thus the susceptibility does not diverge. However, when $T$ approaches $T_{s,M}$ from above, i.e., when we explore the critical behavior for $T\to T_{s,M}^{+}$ even if $M$ jumps suddenly and shows a behavior reminiscent of a first-order transition, the susceptibility $\chi_M$ diverges. This is due to the fact that CE$_{2}$ is the endpoint of a line of second-order PTs between the Para and the Coherent phase. Thus, the magnetization $M$ exhibits the properties of a mixed-order transition.  

In region (iv), for $x_{e,M}<x<x_e'$ we observe a continuous second order PT between the Para and the Coherent phase occurs at $T_{s,M}$. This is due to the behavior of the free-energy densities $f(m)$ and $f(M)$ which display a global minimum at $m=0$ and $M=0$ for $T>T_{s,M}$, while for $T<T_{s,M}$, a global minimum of $f(M)$ appears at finite $M > 0$. The value of this global minimum of $f(M)$ increases continuously as $T$ is lowered. As $T$ is decreased further and reaches $T_f$, new global minima of $f(m_a)$ and $f(M)$ appear at certain finite $m_a$ and $M$. Thus, a first-order transition occurs and both order parameters $m_a$ and $M$ display a discontinuous jump.  The critical behavior of the model in region (iv) is shown in {Fig.~\ref{fig:fig5}}(b) and (f).

At the boundary between the regions (iv) and (v), for $x=x_e'$, we observe the CE$_3$, where the magnetization $m_a$ changes discontinuously from $0$ to a finite value at ${T_s'}^{-}$; however, the susceptibility $\chi_m$ diverges at ${T_s'}^+$. Thus, a mixed-order PT occurs at the CE$_3$.

In region (v), for $x_e'<x<x_g'$ as the temperature is gradually lowered we observe first a send-order PT between the Para and the Coherent phase at $T_{s,M}$, then we observe another second-order PT between the Coherent and the Baxter phase at $T_s'$. In addition to these two PTs we observe a jump of the order parameter $m_a$ and $M$ from non-zero values. This discontinuity can be obtained by studying the free-energy densities $f(m_a)$ and $f(M)$. Indeed when $T > T_{s,M}$,  the global minima of $f(m_a)$ and $f(M)$ remain at $m_a=0$ and $M=0$. For $T_s' < T < T_{s,M}$, the global minimum of $f(M)$ occurs at a finite $M$, which increases continuously as $T$ is lowered gradually. Correspondly, in this same range of temperatures,  the global minimum of $f(m_a)$ remains at still $m_a=0$. As $T$ gets below $T_s'$, a global minimum of $f(m_a)$ emerges at a finite $m_a > 0$ in a gradual way. Thus, $m_a$ is finite, and a second-order  PT occurs at $T_s'$. For this same range of temperature the global minimum of $f(M)$ is achieved at an increasingly larger value of $M$. When $T$ reaches $T_f$, new global minima of $f(m_a)$ and $f(M)$ emerge at finite $m_a$ and $M$, which this minima being separated from the respective value of the free-energy minima obtained for  $T_f^+$. Thus, a discontinuity occurs for the order parameter at $T_f$. These behaviors are schematically shown in {Fig.~\ref{fig:fig5}} (c) and (g). 

In region (vi) corresponding to high values of $x$, or  $x \to \infty$ two second order PTs are observed. The first PT occurs between the Para and the Coherent phase at $T_{s,M}$ and the second PT between the Coherent and the Baxter phase occurs at  $T_s'$. These behaviors close to these two PTs are schematically shown in {Fig.~\ref{fig:fig5}} (d) and (h).

Using Eqs.~\eqref{eq:free_energy_x>1_o} and \eqref{eq:free_energy_x>1_n}, we can obtain the following critical behaviors for $m_a$ and $M$:
\begin{align}
m_a &\sim (T_s'-T)^{\beta_m} && \quad \hbox{with} & ~\beta_m &= \dfrac{1}{\lambda_{\min}-3}, \\
M &\sim (T_{s,M}-T)^{\beta_M} && \quad \hbox{with} & ~\beta_M &= \dfrac{1}{\lambda_\o-3},
\label{eq:beta_exponent_x>1}
\end{align}
where $\lambda_{\min}=\min(\lambda_\o, \lambda_\n)$. Using these results, we obtain the specific heats, which scale as 
\begin{align}
C_m &\sim (T_s'-T)^{-\alpha_m} && \quad \hbox{with} ~~& \alpha_m &= \dfrac{\lambda_\min-5}{\lambda_\min-3} \,, \\
C_M &\sim (T_{s,M}-T)^{-\alpha_M} && \quad \hbox{with} ~~& \alpha_M &= \dfrac{\lambda_\o-5}{\lambda_\o-3} \,. 
\label{eq:alpha_exponent_x>1}
\end{align}
The susceptibilities behave as follows:
\begin{align}
\chi_m \sim 
\left\{ \begin{array}{cccc}
(T-T_s')^{-\gamma_{m}^+} & ~\quad \hbox{with} \quad \gamma_{m}^+ = 1  \,, \\ 
(T_s'-T)^{-\gamma_{m}^-} & ~\quad \hbox{with} \quad \gamma_{m}^- = 1  \,,
\end{array} \right. 
\label{eq:susceptibility_m_>}
\end{align}
and
\begin{align}
\chi_M \sim 
\left\{ \begin{array}{cccc}
(T-T_{s,M})^{-\gamma_{M}^+} & \quad \hbox{with} \quad \gamma_{M}^+ = 1 \,, \\ 
(T_{s,M}-T)^{-\gamma_{M}^-} & \quad \hbox{with} \quad \gamma_{M}^- = 1 \,.
\end{array} \right. 
\label{eq:susceptibility_M_>}
\end{align}
Detailed derivations of $\chi_M$ and $\chi_m$ near $T_{s,M}$ and $T_s'$, respectively, are given in Appendix E.

\begin{table}[h]
\caption{Critical exponents for $3<(\lambda_\n, \lambda_\o)<4$: Here, $\alpha$ is the exponent of the specific heat, $\beta_m$ ($\beta_M$) is the exponent of the magnetization $m_a$ ($M$) at zero external magnetic field, and $\gamma_{m}$ ($\gamma_{M}$) is the exponent of the susceptibility for $m_a$ ($M$)-magnetization near the transition temperature. \label{ta:exponents}}
\begin{normalsize}
\setlength{\tabcolsep}{5pt}
{\renewcommand{\arraystretch}{1.8}
\begin{tabular}{@{\extracolsep{\fill}}cccccccc} 
\hline
\hline
Range of $x$&\multicolumn{1}{c}{$\alpha_m$}&\multicolumn{1}{c}{$\alpha_M$}&\multicolumn{1}{c}{$\beta_m$}&\multicolumn{1}{c}{$\beta_M$}&
\multicolumn{1}{c}{$\gamma_{m\pm}$}& \multicolumn{1}{c}{$\gamma_{M\pm}$} \\
\hline
\hline
$x=0$ &$\frac{\lambda_{\textrm{mim}}-5}{\lambda_{\textrm{min}}-3}$&-&$\frac{1}{\lambda_{\textrm{min}}-3}$&-&$1$&-\\
$0<x<x_M$&$\frac{\lambda_{\min}-5}{\lambda_{\min}-3}$&$\frac{\lambda_{\min}-5}{\lambda_{\min}-3}$&$\frac{1}{\lambda_{\min}-3}$&$\frac{\lambda_\o -2}{\lambda_\o -3}$&$1$&$0$\cr
$x=x_M~(\lambda_\n >\lambda_\o)$&$\frac{\lambda_\o -5}{\lambda_\o -3}$&$\frac{\lambda_\o -5}{\lambda_\o -3}$&$\frac{1}{\lambda_\o -3}$&$\frac{1}{\lambda_\o -3}$&$1$&$1$\cr
$x=x_M~ (\lambda_\n <\lambda_\o)$&$\frac{\lambda_\n -5}{\lambda_\n -3}$&$\frac{\lambda_\o -5}{\lambda_\o -3}$&$\frac{1}{\lambda_\n -3}$&$\frac{1}{\lambda_\o -3}$&$1$&$1$\cr
$x>x_M$ &$\frac{\lambda_{\min}-5}{\lambda_{\min}-3}$&$\frac{\lambda_\o -5}{\lambda_\o -3}$&$\frac{1}{\lambda_{\min}-3}$&$\frac{1}{\lambda_\o -3}$&$1$&$1$\cr
\hline
\hline
\end{tabular}}
\end{normalsize}
\end{table}

\subsection{Anomalous Scaling Relations}

The critical exponents of the continuous transition are listed in Table~\ref{ta:exponents} for all ranges of $x$. 

The scaling relation for $m_a$ satisfies the conventional relation: 
\begin{align}
\alpha_m + 2\beta_m + \gamma_{m} = 2 \,.
\end{align}
By contrast, the scaling relation for $M$ shows an unusual behavior for $x<x_M$. The scaling relation for $M$ does not hold for $x<x_M$ as
\begin{align}
\alpha_M + 2\beta_M + \gamma_{M} = 
\left\{ \begin{array}{cc}
3 &~ \text{for }\lambda_\n >\lambda_\o , \\ \\
3+2\dfrac{\lambda_\o -\lambda_\n}{\lambda_\n -3} &~ \text{for } \lambda_\n <\lambda_\o .
\end{array} \right. \label{eq:scaling_relation}
\end{align}

Note that for the original AT model, the scaling relation for $M$ is written as $\alpha+2\beta_M+\gamma_M=3$. This relation can be confirmed by setting $\lambda_\n=\lambda_\o$ in the second equation of (\ref{eq:scaling_relation}).

\section{Phase diagram II: $\lambda_a$-dependence} 

\subsection{General remarks}

The $g$-AT model may be regarded as a combination of the original AT model on the network of overlapping links and two independent Ising models on the respective network of non-overlapping links. In order to fully appreciate the general phase diagram of the $g$-AT model, let us recall three important results revealing the interplay between network structure and spin models, the Ising, Potts, and AT models.

The Ising model on a single SF network with power-law exponent $\lambda$ exhibits a second-order PT at a finite temperature $T_c \propto \avg{k^2}/\avg{k}$ for $\lambda > 3$ within the annealed approximation~\cite{Ising2}. Thus, as $\lambda$ increases, $T_c$ decreases. Since the magnetization corresponds to the formation of a majority opinion, this implies that the larger the branching ratio $\avg{k^2}/\avg{k}$ of the network is, the easier is for the network to display a majority opinion. In a single SF network, as the power-law {exponent of degree distributions} $\lambda\to 3^+$  the branching ratio of the network increases as $\avg{k^2}/\avg{k}\simeq (\lambda-2)/(\lambda-3)$. Therefore this implies that tuning the power-law exponent $\lambda$, the network undergoes a topological change  that  affects the dynamics of spin model, in particular can modify the value of its critical temperature.

Consequently, we expect that the general phase diagram of the $g$-AT model  will display a significant dependence on the pair of power-law exponents $(\lambda_\n,\lambda_\o)$. In particular the relative value of $\lambda_\n$ with respect to $\lambda_\o$ allows to tune the relative influence of non-overlapping multilinks with respect to overlapping multilinks. We have already seen that $x=J_4/J_2$ modifies the phase diagram as it modulates the strength of the $4$-body interactions (mediated by overlapping multilinks) and the strength of $2$-body interactions (mediated by non-overlapping multilinks). We expect that the phase diagram depends on not only $x$ but also the power-law exponents $(\lambda_\n, \lambda_\o)$ significantly.

Let us recall that the original AT model can be recast in the Potts model with four states when we set $x=J_4/J_2=1$ which display a tricritical PT when the power-law exponent $\lambda_\o=\lambda_c\approx 3.503$. This implies that  for $\lambda_\o > \lambda_c$, the four state Potts model displays a  first-order PT; otherwise, it displays a second-order PT~\cite{Potts}. From this observation we conclude that the phase diagram of the $g$-AT model is expected to be more rich around the values $\lambda_\n\simeq \lambda_c$ and $\lambda_\o\simeq \lambda_c$.

\begin{figure*}
\resizebox{2.0\columnwidth}{!}{\includegraphics{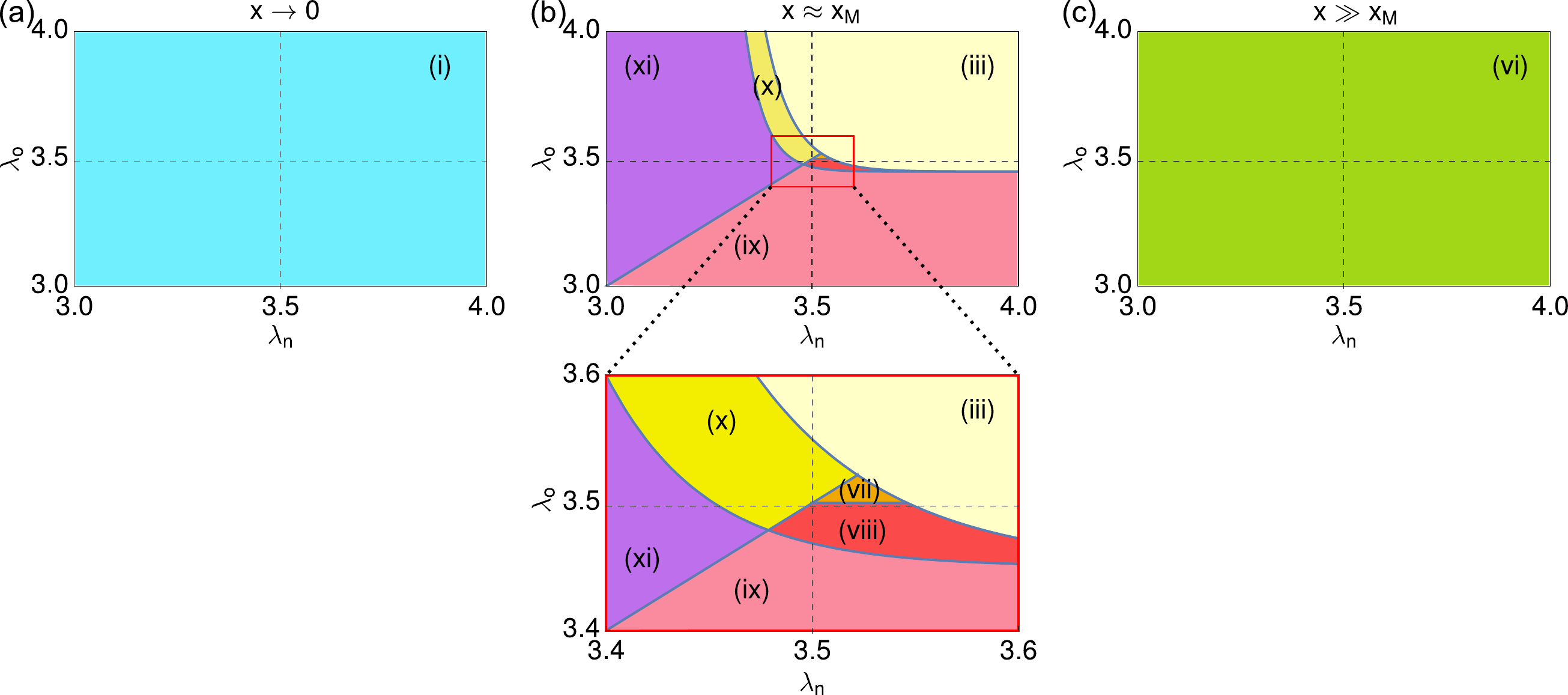}}
\caption{Schematic phase diagrams of the $g$-AT model in the parameter space [$\lambda_\n$, $\lambda_\o$] for (a) $x \approx 0$, (b) $x=x_M$, and (c) $x \gg x_M$. The notations of the phases (i)$-$(vii) are the same as the ones presented in Figs.~\ref{fig:fig3} and \ref{fig:fig10}. \label{fig:fig6}
}
\end{figure*}

\begin{figure*}
\resizebox{2.0\columnwidth}{!}{\includegraphics{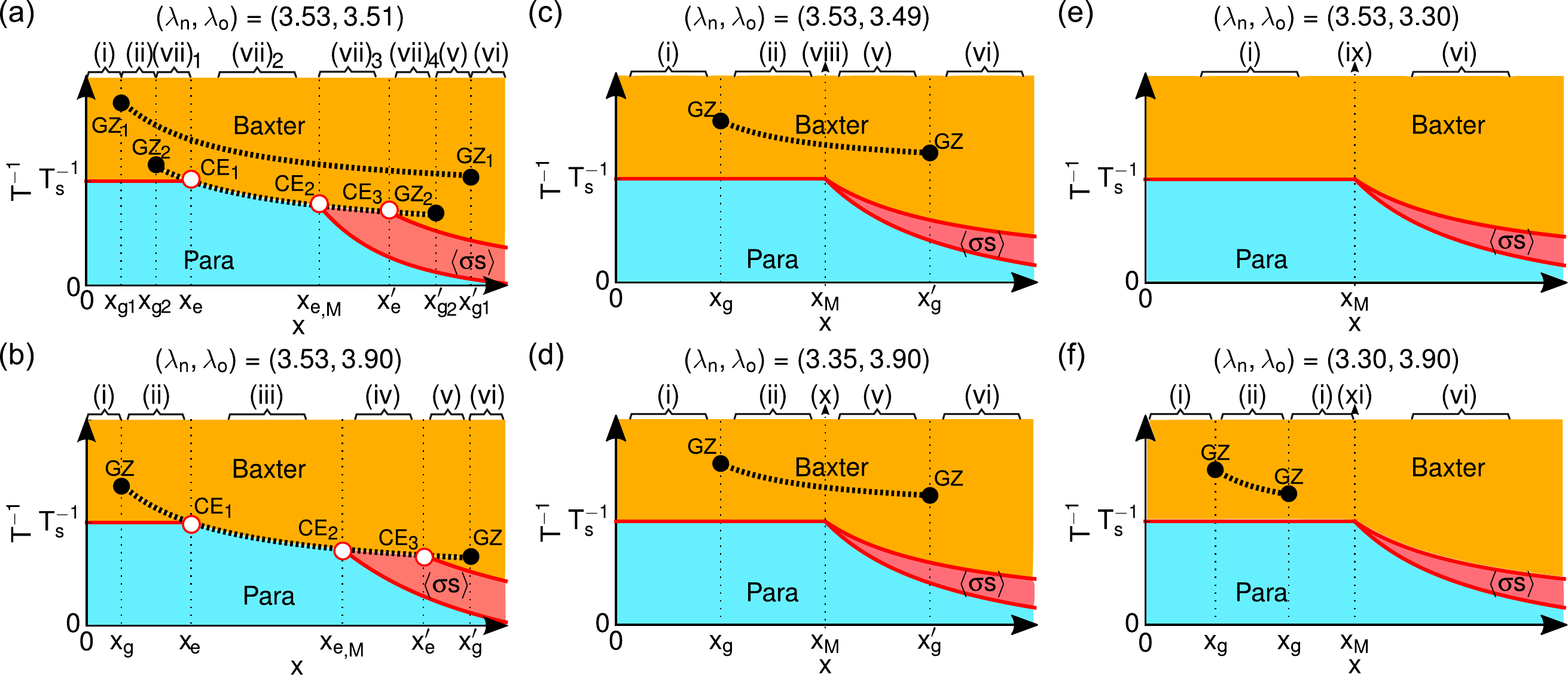}}
\caption{{Schematic phase diagrams of the $g$-AT model in the parameter space [$x, T^{-1}$] for various values $\lambda_\n$ and $\lambda_\o$. ($\lambda_\n$, $\lambda_\o$) = (a) (3.53, 3.51), (b) (3.53, 3.90), (c) (3.53, 3.49), (d) (3.35, 3.90), (e) (3.53, 3.30), and (f) (3.30, 3.90).}
\label{fig:fig7}		
}
\end{figure*}

\subsection{Additional regions of the general phase diagram}

In Sec. IV  we have described the $x$-dependence of the phase diagram of the $g$-AT model  for a choice of power-law exponents   $(\lambda_\n,\lambda_\o) = (3.53,3.90)$. This phase diagram displays the PTs of type (i)-(vi), whose implications for opinion dynamics has been discussed in Sec. IV.

When we consider all possible values of the power-law exponents $(\lambda_\n, \lambda_\o)$ we observe five more characteristic regions and lines, denoted as (vii)--(xi) in Figs.~\ref{fig:fig6} (b):
	 
\begin{itemize}
\item[--] In region (vii) of Fig.~\ref{fig:fig6} and Fig.~\ref{fig:fig7} {(a)}, two discontinuous behaviors occur successively as the temperature is lowered. This region can be divided into four subregions. These subregions are similiar to (ii)-(v) regions in Fig.~\ref{fig:fig7} {(b)} but a discontinuous jump line additionally exists in a lower temperature region. The discontinuous jump line originates from the correlations between $m_\n$ of one layer and $m_\n$ or $m_\o$ of the other layer, whereas the other discontinuous jump lines in the interval $[x_{g2},x_e]$ and $[x_e',x_{g1}']$ in a higher temperature region originates from the correlation between the same $m_\o$s but on different layers.
\item[1.]In region (vii)$_1$ ($x_{g2} < x < x_{e}$), a continuous PT between the Para and the Baxter phase occurs at $T_s$ and two discontinuous jumps of the order parameters $m_a$ and $M$ occur successively as the temperature is lowered. As the noise (temperature) is reduced, the opinion dynamics exhibits first a continuous PT in which a majority opinion is formed in both layers, and spreads abruptly twice over a finite fraction of nodes in the multiplex network.
\item[2.]
In region (vii)$_2$ ($x_{e}<x<x_{e,M}$), as the temperature is lowered, a discontinuous PT between the Para and the Baxter phase occurs firstly and subsequently a discontinuous jump of the order parameters $m_a$ and $M$ occurs in the same Baxter phase. Therefore as the noise is reduced, the opinion dynamics has first a discontinuous PT in which a majority opinion is formed in both layers, and {then we
observe an additional jump in the magnetization as the majority opinion gets adopted by a larger fraction of nodes of the multiplex network.}
\item[3.]
In region (vii)$_3$ ($x_{e,M}<x<x_{e}'$), a continuous PT occurs between the Para and the Coherent  phase at $T_{s,M}$. As the temperature is decreased further, a discontinuous PT occurs at $T_{f}$ from the Coherent phase to the Baxter phase and a {discontinuous jump} of the order parameters $m_a$ and $M$ occurs at $T_{f}'$ successively. This implies that as the noise is reduced, at temperatures {lower than} $T_{s,M}$, each single {node prefers} to adopt a coherent opinion in both layers. As the temperature is further reduced, {we observe a discontinuous PT in which majority opinion is formed in both layers, and then the majority opinion spreads abruptly over a finite fraction of nodes in the multiplex network.}
\item[4.]
In region (vii)$_4$ ($x_{e}'<x<x_{g2}'$), two continuous PTs occur successively: between the Para and the Coherent phase at $T_{s,M}$ and between the Coherent and the Baxter phase at $T_s'$. Then as the temperature is decreased further two discontinuous jumps of the order parameter occur at $T_{f}$ and $T_{f}'$, respectively. Overall, as the noise is reduced, the opinion dynamics exhibits two {types of} continuous PTs {successively} in which the coherent and the majority opinion {are} formed, respectively. {When} the noise is decreased further, the majority opinion spreads abruptly twice over {a finite fraction of nodes} in the multiplex network. 
\item[--] In regions or points (viii) and (x) of Figs.~\ref{fig:fig6} and Fig.~\ref{fig:fig7} {(c) and (d)}, respectively, a continuous PT between the Para and the Baxter phase and a discontinuous jump of the order parameters $m_a$ and $M$ occur successively as the temperature is lowered. The critical behavior at $x_M$ differs from that in region (ii) as we observe at a tricritical point. The transition point at $x_M$ in (viii) and (x) acts as a branching point of the critical line to the two critical lines of the Coherent phase. Overall, as the noise is reduced the opinion dynamics exhibits first a continuous PT in which a majority opinion is formed in both layers, and subsequently the majority opinion spreads abruptly over a large fraction of nodes in the multiplex network.
\item[--] In regions (ix) and (xi) of Fig.~\ref{fig:fig6} and in Fig.~\ref{fig:fig7} {(e) and (f)} at $x_M$, respectively, a continuous PT between the Para and the Baxter phase is observed as the temperature is lowered. The critical behavior at $x_M$ differs from that in (i) as we observe at a tricritical point. The transition point at $x_M$ in (ix) and (xi) acts as a branching point of the critical line to the two critical lines of the Coherent phase. As the noise is reduced, the opinion dynamics has continuous PT in which a majority opinion is formed in both layers.
\end{itemize}



In order to explore the dependence of this rich phase diagram on the exponents $\lambda_\n$ and $\lambda_\o$ around $x=x_M$, we plot the phase diagram in the space [$x, T^{-1}$] for various values of the degree pairs $(\lambda_\n, \lambda_\o)$ (see Fig.~\ref{fig:fig7}). We find that if a first-order PT occurs at $x=x_M$, the overall phase diagram is close to the phase diagram discussed in Sec. IV (see Fig. $\ref{fig:fig3}$ and Fig.~\ref{fig:fig7} (b)). If $x_M$ demarks the boundary between type-(ii) and type-(v) PTs, then the phase diagram is similar to Fig.~\ref{fig:fig7} (c) and (d), respectively. When $\lambda_\n < \lambda_\o \approx 3.90$ (see Fig.~\ref{fig:fig7} (b), (d), and (f)), as $\lambda_\n$ is decreased, a discontinuous transition curve shrinks and moves left and upward as shown in Figs.~\ref{fig:fig7} (b)$\to$(f). Moreover, when $\lambda_\n$ is slightly larger than $\lambda_\o \approx \lambda_c^+$ (see Fig.~\ref{fig:fig7} (a)), double discontinuous transition curves appear in the phase diagram, where two discontinuous PTs occur successively as $T$ is decreased.

\subsection{Free-energy landscape for the $\lambda$-dependence of phase transitions at $x \approx x_M$}

Here, we will investigate the free-energy landscape of the $g$-AT model for the (vii)--(xi) types of PTs. Around $x \approx x_M$, phase and PT type are determined by the free energy density presented in Appendix D. 

In region (vii)$_1$, PT type is determined by the free energy density given as Eq.~\eqref{eq:free_energy_x<1_o} because $x < x_M$. Note that the higher-order term \eqref{eq:additional_x<1_o} of Eq.~\eqref{eq:free_energy_x<1_o} is negative. As $x \to x_M$, the terms with $B_\o$ of Eq.~\eqref{eq:additional_x<1_o} and with $D_m$ of Eq.~\eqref{eq:free_energy_x<1_o} become comparable in their magnitudes to the terms with $C_3$ and $D_\o$ in Eq.~\eqref{eq:free_energy_x=1_II}, respectively. Thus, these terms with $D_m$ and $B_\o$ play a similar role to the terms with $C_\o$ and $C_3$. For $T>T_s$, the terms with $D_m$ and $B_\o$ are not large in magnitude, so that the global minimum of $f(m_a)$ remains at $m_a=0$ and $M=0$, and thus a continuous PT occurs at $T_s$. However, when $T$ is lowered further, the term with $D_m$ increases and becomes comparable to the leading order terms. Then a discontinuous {jump of the order} parameters occurs at $T_f$. As $T$ is lowered further, another negative term with $B_\o$ term increases, another jump of the order parameters occurs at $T_f'$. Hence in the region (vii)$_1$, as $T$ is lowered from $T_s^+$, a second-order PT occurs firstly and then two discontinuous jumps occur successively in the Baxter phase.

\begin{figure}
\resizebox{1.0\columnwidth}{!}{\includegraphics{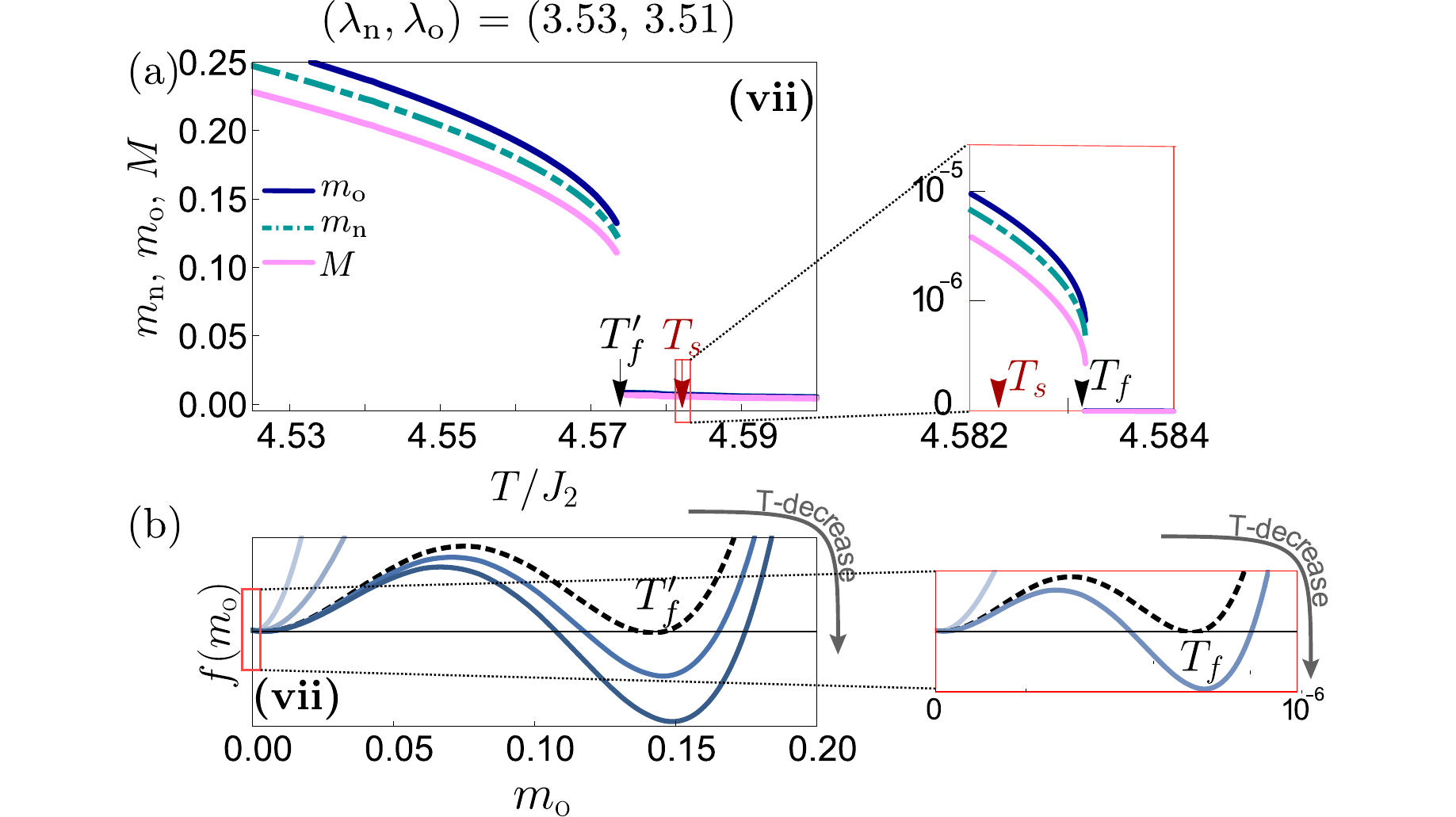}}
\caption{For the (vii)-type of PT, schematic plots of (a) the order parameters $m_a$ and $M$ as a function of $T$ and (b) the free energy {density} landscape as a function of $m_\o$ for various $T$s. The {exponents of degree distributions} are taken as ($\lambda_\n$, $\lambda_\o$) = (3.53, 3.51).
\label{fig:fig8}	
} 	
\end{figure}

In region (vii)$_2$, the term with $C_\o$, induced by the correlation between $m_\o$s on different layers, becomes negative when $\lambda_\o > \lambda_c$, thus produces a discontinuous PT at $T_f$ higher than $T_s$. Another negative term with $C_3$, induced by the correlations between $m_\n$ of one layer and $m_\n$ or $m_\o$ of the other layer, becomes larger as $T$ is decreased, and thus a discontinuous jump of the order parameters $m_a$ and $M$ in the Baxter phase occurs at $T_f'<T_s$. $f(m_a)$ and $f(M)$ develop a global minimum at a temperature $T_{f}$, leading to a discontinuous PT between the Para and the Baxter phase. As $T$ is further lowered from $T_{f}$, the global minimum position of $m_a$ and $M$ increases continuously until a certain temperature $T_{f}'$. When $T$ reaches $T_{f}'$, another global minimum of $f(m_a)$ and $f(M)$ emerge at another finite $m_a$ and $M$, which lead to the jumps in the order parameters $m_a$ and $M$ in the Baxter phase. The order parameters and free energy landscape in this region are depicted in Fig.~\ref{fig:fig8} (a) and (b), respectively. 

In region (vii)$_3$, PT type is determined by the free energy density  Eq.~\eqref{eq:free_energy_x>1_o}, because $x > x_M$. Note that the higher-order term \eqref{eq:additional_x>1_o} of Eq.~\eqref{eq:free_energy_x>1_o} is negative. As $x \to x_M$, the terms with $B_\o'$ of Eq.~\eqref{eq:additional_x>1_o} and with $D_M$ of Eq.~\eqref{eq:free_energy_x>1_o} become comparable in their magnitudes to the terms with $C_3$ and $D_\o$ in Eq.~\eqref{eq:free_energy_x=1_II}, respectively. Thus, these terms with $D_M$ and $B_\o'$ play a similar role to the terms with $C_\o$ and $C_3$. For $T>T_{s,M}$, the terms with $D_M$ and $B_\o'$ are small in magnitude, so that a global minimum of $f(m_a)$ and $f(M)$ remain at $m_a=0$ and $M=0$, respectively. As $T$ is lowered from $T_{s,M}^+$, a second-order PT for $M$ occurs from the Para to the Coherent phase at $T_{s,M}$ and a global minimum of $f(M)$ increases continuously. When $T$ is decreased further to $T_f$, the term with $D_M$ increases in its magnitude, a new global minimum of $f(m_a)$ and $f(M)$ {appears far from} values at $T_f^+$, respectively, a discontinuous PT between the Coherent and the Baxter phase appears at $T_f$. When $T$ is lowered further, the term with $B_\o'$ becomes large, a discontinuous jump of the order parameters occurs at $T_f'$ in the Baxter phase. Hence, in the region (vii)$_3$, as $T$ is lowered from $T_{s,M}^{+}$, a continuous PT from the Para to the Coherent phase occurs firstly at $T_{s,M}$ and then a discontinuous PT between the Coherent and the Baxter phase occurs at $T_f$ and then a discontinuous jump occurs at $T_f'$ successively.

In region (vii)$_4$, PT type is investigated through   Eq.~\eqref{eq:free_energy_x>1_o}, because $x > x_M$. When $T > T_{s,M}$, the terms with $D_M$ and $B_\o'$ are too small, and the global minimum of $f(m_a)$ and $f(M)$ remain at $m_a=0$ and $M=0$, respectively. For $T_s' < T < T_{s,M}$, a second-order PT for $M$ from the Para to the Coherent phase occurs at $T_{s,M}$, and the global minimum of $f(M)$ grows continuously as $T$ is lowered from $T_{s,M}^+$. Meanwhile, the global minimum of $f(m_a)$ remains at still $m_a=0$. As $T$ is lowered across $T_s'$, a second-order PT for $m_a$ from the Coherent to the Baxter phase occurs at $T_s'$. When $T$ is lowered and reaches $T_f$ and $T_f'(<T_f)$, the global minimum of $f(m_a)$ and $f(M)$ jump discontinuously from the previous positions at $T_f^+$ and $T_f^{'+}$, respectively. Hence, in the region (vii)$_4$, as $T$ is decreased from $T_{s,M}^{+}$, a continuous PT from the Para to the Coherent phase occurs at $T_{s,M}$ and a continuous PT from the Coherent to the Baxter phase occurs at $T_s'$, and then two discontinuous jump of the order parameters in the Baxter phase occur at $T_f$ and $T_f'$, successively.

\begin{figure}
\resizebox{1.0\columnwidth}{!}{\includegraphics{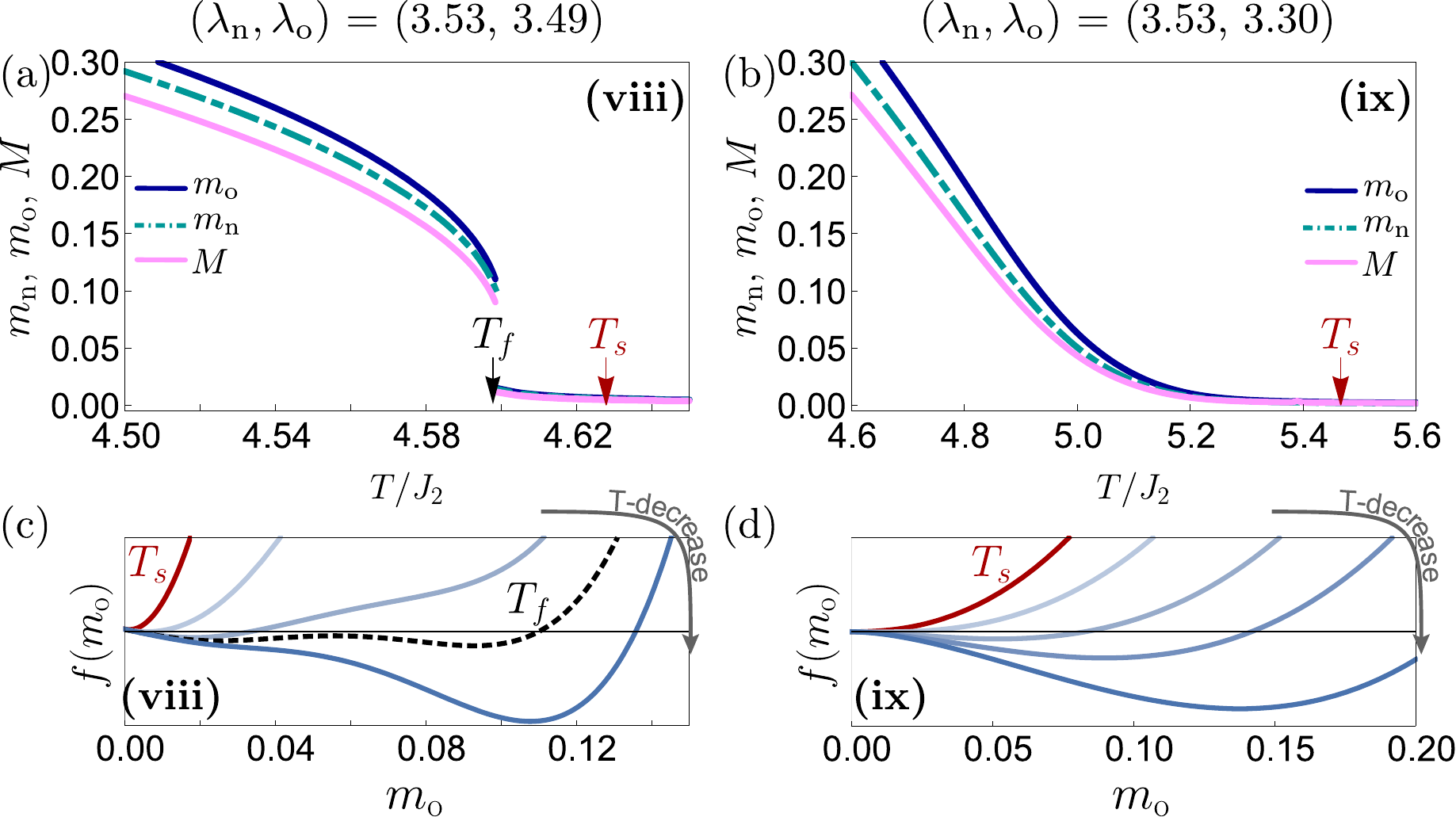}}
\caption{For the (viii)- and (ix)-type of PTs, schematic plots of (a) and (b) the order parameters $m_a$ and $M$ as a function of $T$, respectively. (c) and (d) schematic plots of the free energy {density} landscape as a function of $m_\o$ for the temperatures around $T_f$. The {exponents of degree distributions} for (a) and (c) are taken as ($\lambda_\n$, $\lambda_\o$) = (3.53, 3.30) and for (b) and (d) are taken as (3.53, 3.49).
\label{fig:fig9}		
}
\end{figure}

In region (viii), $f(m_a)$ and $f(M)$ still display a global minimum at $m=0$ and $M=0$ for $T>T_s$ and a continuous PT between the Para and the Baxter phase at $T=T_s$. The values of the critical exponents for this PT are listed in Table I for the case $x\approx x_M$ and $\lambda_\n > \lambda_\o$. However, as $T$ is further lowered below a certain temperature $T_f$, a global minimum emerges at a non-zero value of the magnetization $m_a > 0$ and $M>0$, and a discontinuous transition occurs. Thus, as $T$ is decreased from $T_s^+$, a continuous transition occurs firstly at $T_s$, followed by a discontinuous transition at $T_f$. The order parameters and free energy density landscape are depicted in Fig.~\ref{fig:fig9} (a) and (c), respectively. 

In region (ix), $f(m_a)$ and $f(M)$ have a global minimum at $m_a=0$ and $M=0$, respectively, for $T \ge T_s$ while for  $T=T_s^{-}$ a global minimum emerges continuously at a non-zero value of the magnetization. Therefore at $T=T_s$ we observe a continuous PT between the Para and the Baxter phase. The values of the critical exponents for this PT are listed in Table I for the case $x\approx x_M$ and $\lambda_\n > \lambda_\o$. The order parameters and free energy density landscape are illustrated in Fig.~\ref{fig:fig9} (b) and (d), respectively. This continuous transitions is similar to the (i)-type PTs; however, the critical behavior for $M$ of this continuous transition differs from that of the (i)-type PT, and thus we denote this type of a continuous transition as the (ix)-type PT to distinguish this from the (i)-type PTs.

\begin{figure}
\resizebox{1.0\columnwidth}{!}{\includegraphics{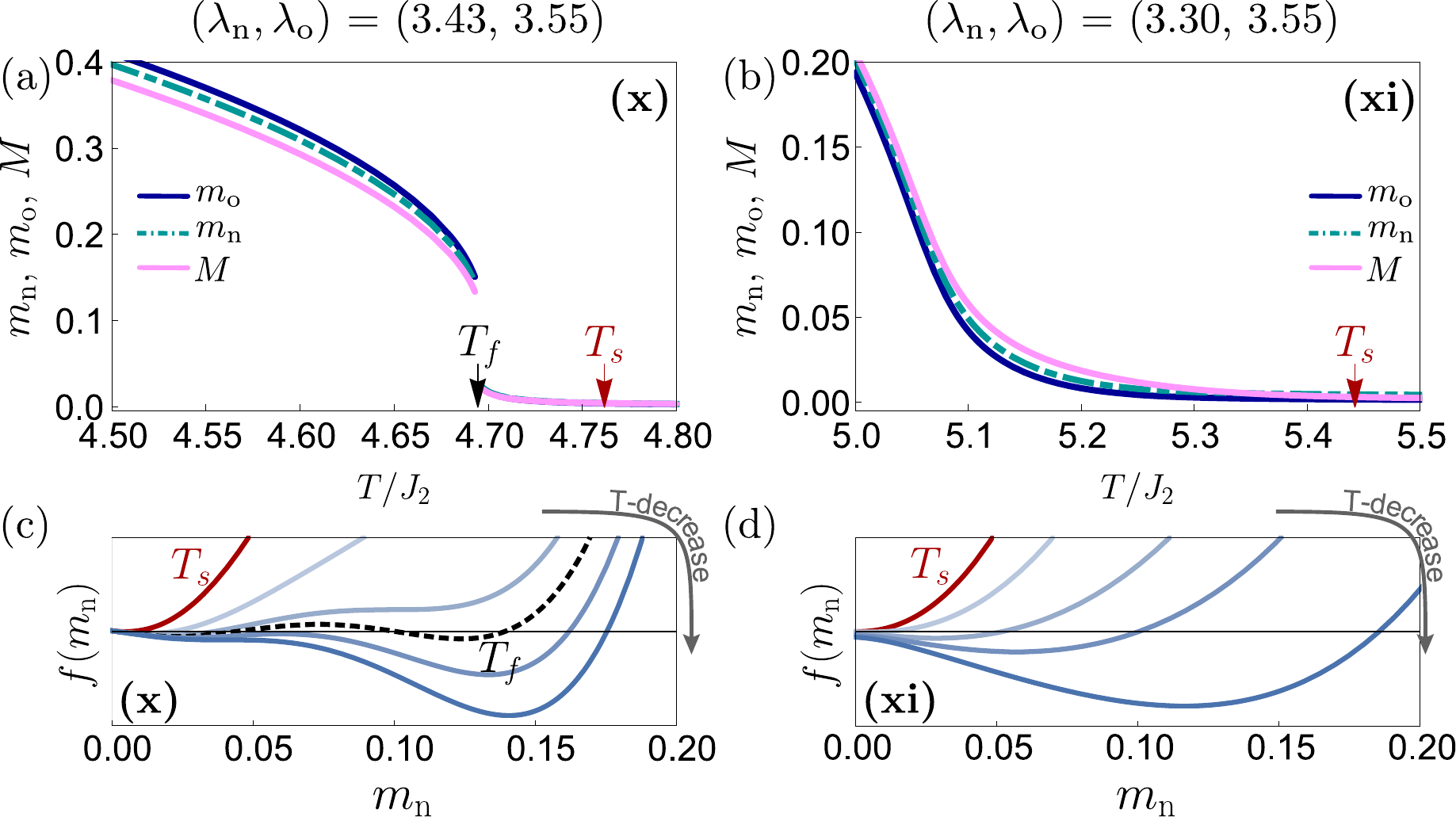}}
\caption{(a) and (b) Plots of the order parameters $m_a$ and $M$ at $x=x_M$ as a function of $T/J_2$. (c) and (d) Plots of the free energy {density} landscape as a function of $m_\n$ for various $T$. The {exponents of degree distributions} taken for (a) and (c) are ($\lambda_\n$, $\lambda_\o$) = (3.30, 3.55); for (b) and (d) are (3.43, 3.55).	
\label{fig:fig10}	
} 	
\end{figure}




In region (x), $f(m_a)$ and $f(M)$ have a global minimum at $m_a=0$ and $M=0$, respectively. $f(m_a)$ remains at $m_a=M=0$ for $T>T_s$. When $T$ is lower than $T_s$, a continuous transition occurs. The values of the critical exponents for this PT are listed in Table I for the case $x\approx x_M$ and $\lambda_\n < \lambda_\o$. As $T$ is further lowered, the order parameter gradually increases. When $T$ reaches $T_f$, the order parameter jumps by a finite amount and a new global minimum of $f(m_a)$ occurs at a finite $m_a$. Thus, as $T$ is decreased from $T_s^+$, a continuous PT occurs at $T_s$ firstly and then a discontinuous jump of the order parameter occurs at $T_f$ as shown in Fig.~\ref{fig:fig10}(a) and (b). 

In region (xi), $f(m_a)$ and $f(M)$ have a global minimum at $m_a=0$ and $M=0$, respectively. They remain at $m_a=M=0$ for $T>T_s$.
At $T_s$, $m_a$ and $M$ exhibit continuous PTs. The values of the critical exponents for this PT are listed in Table I for the case $x\approx x_M$ and $\lambda_\n < \lambda_\o$. When $T$ is decreased from $T_s$, a global minimum occurs at finite $m_a$ and $M$. These behaviors are schematically shown in Fig.~\ref{fig:fig10} (b) and (d). Note that the $\beta_M$ and $\chi_M$ of the (xi)-type PTs are different from those of the (i)-type PTs (Eq.~\eqref{eq:beta_exponent_x<1} and $\gamma_M=0$).

\begin{figure}
\resizebox{0.85\columnwidth}{!}{\includegraphics{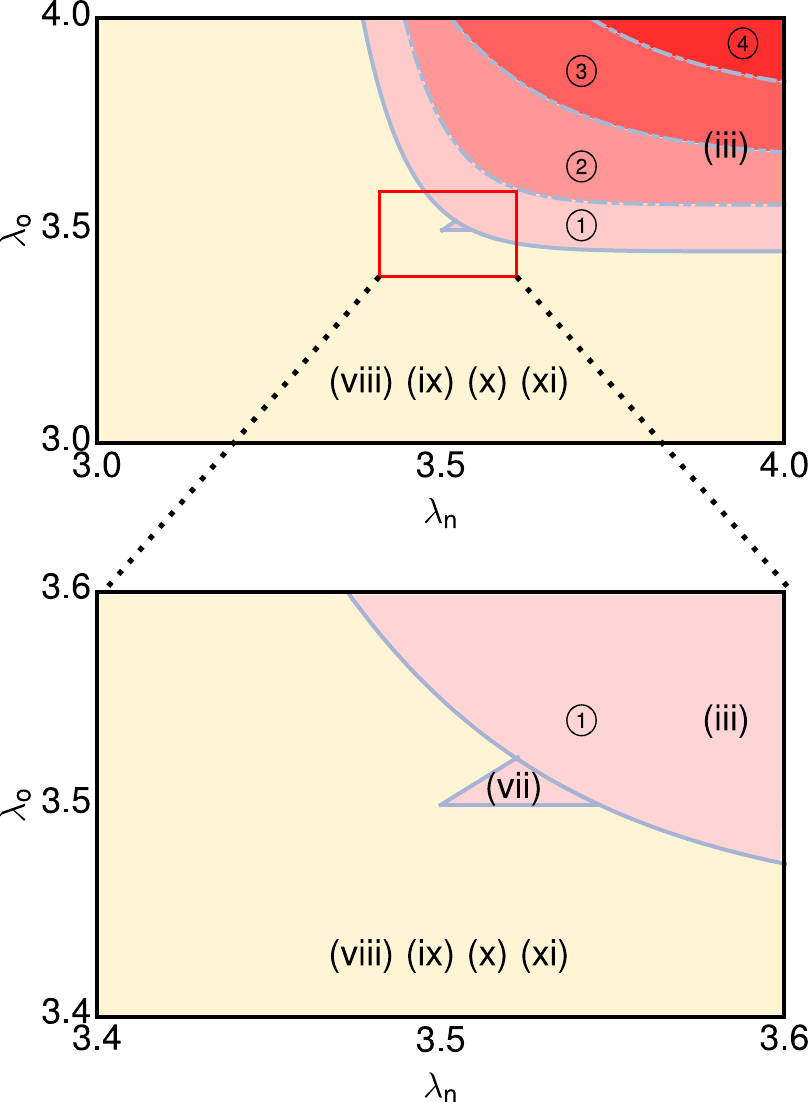}}
\caption{Schematic contour lines of $x_{e,M}$ in the parameter space [$\lambda_\n$, $\lambda_\o$]. The rightmost (light yellow) region contains the domains (viii), (ix), (x) and (xi) as shown in Fig.~\ref{fig:fig6} (b). In this region, a continuous PT appears at $T_s$ at $x_M$, and thus $x_{e,M} = x_M$, the Coherent phase appears at $x_M$. The regions denoted as {\textcircled{\scriptsize 1}}$-${\textcircled{\scriptsize 4}} correspond to the regions (iii) and (vii) denoted in Fig.~\ref{fig:fig6} (b). In this region, a discontinuous PT appears at $x_M$, and thus $x_{e,M} > x_M$. The contour lines represent in term of the ratio $x_{e,M}/x_M$. 
In {\textcircled{\scriptsize 1}}, $1 < x_{e,M}/x_M < 1.05$; in  {\textcircled{\scriptsize 2}}, $1.05 < x_{e,M}/x_M <1.10$; in  {\textcircled{\scriptsize 3}}, $1.10 <x_{e,M}/x_M < 1.15$; and in {\textcircled{\scriptsize 4}}, $1.15 < x_{e,M}/x_M$.
\label{fig:fig11}
}
\end{figure}

\subsection{$\lambda$-dependence of the Coherent phase}

When $x \to 0$, a second-order PT occurs from the Baxter to the Para phase, whereas when $x \gg x_M$, a first-order PT occurs from the Baxter to the Coherent phase (i.e., $\langle \sigma s \rangle$), followed by another PT occurs from the Coherent to the Para phase. The Coherent phase appears for $x \ge x_{e,M}(\lambda_a)$. In Fig.~\ref{fig:fig11},  we display the $\lambda_a$-dependence of $x_{e,M}$ in unit of $x_M$. If $x_{e,M}=x_M^+$, then the Coherent phase appears in the range of $x > x_M$ as shown in Fig.~\ref{fig:fig7} (a), (b), (d), and (e), which is denoted as (ix), (viii), (xi), and (x) (light yellow) in Fig.~\ref{fig:fig11}, respectively. In the regions (iii) and (vii), when a discontinuous PT occurs at $x_M$, then $x_{e,M} > x_M$, and the Coherent phase appears in range of $x > x_{e,M}$ as shown in Fig.~\ref{fig:fig7} (c) and (f), respectively. The contour lines between different regions with different circle numbers represent different ratios $x_{e,M}/x_{M}$.

We find that the transition point $x_{e,M}$ is delayed as both $\lambda_{\n}$ and $\lambda_{\o}$ are increased. This is caused by the following reasons: if $\lambda_{\n}$ and $\lambda_{\o}$ are large, then the branching ratios of non-ovelapping and overlapping links, respectively, become small. Thus, a larger value of $x=J_4/J_2$ is needed, i.e., the strength of 4-body interaction needs to be reinforced to form a Coherence phase. 

\section{Conclusion}

To investigate the effect of link overlap on the opinion dynamics  defined on a multiplex network, we studied the so-called $g$-AT model, a spin model in thermal equilibrium systems. The $g$-AT model describes the dynamics of two species of Ising spins, namely the $s$ and $\sigma$ spins, each of which is located on a single layer of the duplex network under consideration. Here, the spin model is defined on duplex networks with an SF multidegree distribution, which facilitates tuning of the effect of overlapping links with respect to non-overlapping links. In particular, we distinguish between multilinks $(1,1)$ that characterize overlapping links and multilinks $(1,0)$ and $(0,1)$ that do not. We assume that the multidegrees $k^{(1,1)}=k_\o$ and $k^{(1,0)}=k^{(0,1)}=k_\n$ follow power-law distributions associated with the tunable power-law exponents $\lambda_\n$ and $\lambda_\o$. This system is illustrated in Fig.~\ref{fig:1a}.

Pairs of $s$-spins (pairs of $\sigma$-spins) connected by overlapping and non-overlapping links interact through a 2-body interaction of strength $J_2$. Four spins comprising two $s$-spins and two $\sigma$-spins connected by overlapping links interact through a 4-body interaction with strength $J_4$ (see Fig. \ref{fig:1b}). The ratio $x\equiv J_4/J_2$ is a control parameter that can alter the critical properties of the model, and the system is assumed to be in thermal contact with a heat reservoir at temperature $T$. Here, $T$ represents the diversity of each individual opinion in a social community. Thus, there exist four control parameters, namely $\lambda_\n$, $\lambda_\o$, $x$, and $T$. By applying the Landau--Ginzburg theory, we obtained rich phase diagrams in the four-parameter space. The $g$-AT model is a generalization of the original AT model~\cite{AT}, in which all the links are regarded as overlapping links; therefore, a single exponent $\lambda_\o$ is considered in this context. 

We note that the different species of spins represent individuals from two different communities formed based on friendship and business relations, respectively. Each pair of individuals may be connected solely via friendship links, solely via business relations, or via both relationships. The formation of a majority opinion across both layer is indicated by the magnetizations $\langle \sigma \rangle>0$, $\langle s \rangle>0$, and $\langle \sigma s \rangle>0$ in the spin model, which can be accomplished through non-overlapping or overlapping links. The diversity of individual opinions is reflected by thermal fluctuations. 

We investigated PTs arising from the competition between the consensus formation of each community and that of the entire society, and obtained rich phase diagrams including diverse types of PTs. These findings are expected to be beneficial in understanding the underlying mechanisms of local and global  formation of a majority opinion in a society. 

Similar to the voter models on multiplex networks~\cite{Chmiel1}, the $g$-AT model shows that a majority opinion emerges abruptly thanks to the interactions across two layers induced by overlapping links. In particular, in the $g$-AT model, we can control the strength of 4-body interactions among replica nodes connected by overlapping multilinks with respect of the strenght of 2-body interactions $x=J_4/J_2$. This allows us to assess the role of  tuning the strength of the many-body AT-interactions by modulating $x$ and study how the phase digram change with respect to the original AT model ~\cite{AT}. Moreover we can tune the power-law exponent of the  overlapping multidegree distributions of non-overlapping and overlapping multilinks $(\lambda_\n, \lambda_\o)$, and investigate the role of this topological modifications on PTs, in the same spirit of the analysis conducted for percolation problems in Ref.~\cite{Cellai1, Cellai2}. In the future our work can be expanded in many directions, investigating further the role that higher-order interactions \cite{battiston,HO} have in opinion dynamics defined on multiplex networks and exploring  realistic spin models of opinion dynamics defined on duplex networks.

\appendix

\begin{widetext} 
\section{Self-consistency equation with an external magnetic field}
With the {external magnetic field}
\begin{align}
\Omega_a \equiv K_2 m_a  + H_a, \quad \textrm{and} \quad \Omega_M \equiv K_4 M  +H_4,
\end{align}
the self-consistency equations Eqs.~\eqref{eq:equationOfState_ext_field} for {$m_a$} and $M$ of the $g$-AT model are replaced as follows:
\begin{align}
	m_a \langle k_a \rangle 
	&= m_{a\I} \equiv \int_{k_{\rm min}^\n}^{\infty} \int_{k_{\rm min}^\o}^{\infty} dk_\n dk_\o  P_d(k_\n) P_d(k_\o) \, \dfrac{\tanh\left(\Omega_\o k_\o + \Omega_\n k_\n \right)\left[1+\tanh\left(\Omega_4 k_\o \right)\right]}{1+\tanh^2\left(\Omega_\o k_\o + \Omega_\n k_\n \right) \tanh\left(\Omega_4 k_\o \right)} k_a
	\label{eq:equationOfState_m_H}
\end{align}
and
\begin{align}
M\langle k_\o  \rangle
&= M_\I \equiv \int_{k_{\rm min}^\n}^{\infty} \int_{k_{\rm min}^\o}^{\infty} dk_\n dk_\o  P_d(k_\n) P_d(k_\o) \dfrac{\tanh \left(\Omega_4 k_\o \right) +\tanh^2 \left(\Omega_\o k_\o + \Omega_\n k_\n \right)}{1 + \tanh^2 \left(\Omega_\o k_\o + \Omega_\n k_\n \right) \tanh \left(\Omega_4 k_\o \right)} k_\o \,.
\label{eq:equationOfState_M_H}
\end{align}

\section{Definitions of the $\mathcal{A}$ terms in exact susceptibility formula}
The $\A$ terms are defined as follows:

\begin{align}
	\A_{aa} 
	&= \int_{k_{\rm min}^\n }^{\infty} \int_{k_{\rm min}^\o }^{\infty} dk_\n  dk_\o  P_d(k_\n) P_d(k_\o) \dfrac{(1-\T_2^2 \T_4)(1+\T_4)}{(1+\T_2^2 \T_4)^2 \cosh^2\left(K_2 (m_\o k_\o  + m_\n k_\n) \right)} k_ a  ^2  \,, \\ \cr
	\A_{a \bar a  } = \A_{\bar a   a} 
	&= \int_{k_{\rm min}^\n }^{\infty} \int_{k_{\rm min}^\o }^{\infty} dk_\n  dk_\o  P_d(k_\n) P_d(k_\o) \dfrac{(1-\T_2^2 \T_4)(1+\T_4)}{(1+\T_2^2 \T_4)^2 \cosh^2\left(K_2 (m_\o k_\o  + m_\n k_\n) \right)} k_\n k_\o \,, \\ \cr
	\A_{a M}
	&= \int_{k_{\rm min}^\n }^{\infty} \int_{k_{\rm min}^\o }^{\infty} dk_\n  dk_\o  P_d(k_\n) P_d(k_\o) \dfrac{(1-\T_2^2) \T_4}{(1+\T_2^2 \T_4 )^2 \cosh^2 (K_4 M k_\o)} k_a  k_\o \,, \\ \cr
	\A_{M a}
	&= \int_{k_{\rm min}^\n }^{\infty} \int_{k_{\rm min}^\o }^{\infty} dk_\n  dk_\o  P_d(k_\n) P_d(k_\o) \dfrac{2\T_2 (1-\T_4^2 )}{(1+\T_2^2 \T_4 )^2 \cosh^2 \left(K_2 (m_\o k_\o  + m_\n k_\n) \right)} k_\o k_a \,, \\ \cr
	\A_{M M} 
	&=  \int_{k_{\rm min}^\n }^{\infty} \int_{k_{\rm min}^\o }^{\infty} dk_\n  dk_\o  P_d(k_\n) P_d(k_\o) \dfrac{1-\T_2^4 }{(1+\T_2^2 \T_4 )^2 \cosh^2 (K_4 M k_\o)} k_\o^2 \,. 
\end{align}

\section{Definitions of coefficients in the free energy density}
\label{seca:A}
The coefficients $C_\o$, $C_\n$, $D_m$, $D_M$, $D_0$, and $C_\o(r_0, \lambda_\o)$ used in the Landau free energy formulas in Appendix D are defined as follows:
\begin{align}
C_\o  (\lambda_\o) 
& =  C_{M} (\lambda_\o) = - N_\o  \int_0^\infty \left[\ln (\cosh y) - \frac{1}{2} y^2 \right] y^{-\lambda_\o} dy, \cr
C_\n  (\lambda_\n)
& = - N_\n  \int_0^\infty \left[\ln (\cosh y) - \frac{1}{2} y^2 \right] y^{-\lambda_\n} dy, \cr
D_m (\lambda_\o)
& = - N_\o  \int_0^\infty y \ln \left( 1+\tanh^2 y \right) y^{-\lambda_\o} dy, \cr
D_M (\lambda_\o)
& = - N_\o  \int_0^\infty y^2 \ln \left(1+\tanh y \right) y^{-\lambda_\o} dy,\cr
D_0 (r_0, \lambda_\o) 
& = - N_\o  \int_0^\infty \ln \left(1+\tanh^2 y \tanh(r_0 y) \right) y^{-\lambda_\o} dy, \cr
C_\o(r_0, \lambda_\o)
& = C_\o (\lambda_\o) + D_0(r_0,\lambda_\o)+ \frac{1}{2}C_{M}(\lambda_\o),
\end{align}
where $N_\o$ and $N_\n$ are normalization factors written as ${1}/({\lambda_\o-1})$ and ${1}/({\lambda_\n-1})$, respectively.

\section{Landau Free Energy Formula}

To investigate the critical behavior near the critical temperature, we expand the free energy density as a function of the order parameters $m_a$ and then analyze the leading terms when $m_a$ and $M$ converge to $0$. To proceed, it is necessary to derive  the relation between $m_a$ and $M$, which turns out to depend on the ratio $x$. For values of $x$ smaller and bigger than the characteristic ratio $x_M$, we observe different behaviors. Here we discuss in details the cases $0< x < x_M$, $x > x_M$, and $x=x_M$, separately. \\

\subsection{Case $ 0 < x < x_M$} 

For $x\in (0,x_M)$ we can expand Eqs.~\eqref{eq:equationOfState_m} and~\eqref{eq:equationOfState_M}  in terms of $m_a$ and $M$ within the lowest order terms as follows: 
\begin{align}
m_\n \langle k_\n \rangle \left(1- K_2 \dfrac{\langle k_\n^2 \rangle}{\langle k_\n \rangle}\right)
& \simeq K_2 m_\o \langle k_\n \rangle \langle k_\o \rangle - (\lambda_\n-1)C_\n (\lambda_\n)(K_2 m_\n)^{\lambda_\n-2} + \hbox{h.o.,} \label{eq:small_mn} \\ \cr
m_\o \langle k_\o \rangle \left(1- K_2 \dfrac{\langle k_\o^2 \rangle}{\langle k_\o \rangle}\right) 
& \simeq K_2 m_\n \langle k_\n \rangle \langle k_\o \rangle- (\lambda_\o-1)C_\o (\lambda_\o)(K_2 m_\o)^{\lambda_\o-2} + (\lambda_\o-2)D_{m}(\lambda_\o)(K_4M)(K_2 m_\o)^{\lambda_\o -3} + \hbox{h.o.,}
\label{eq:small_mo} \\ \cr
M \langle k_\o \rangle \left(1- x K_2 \dfrac{\langle k_\o^2 \rangle}{\langle k_\o \rangle}\right) 
&\simeq D_m(\lambda_\o)(K_2 m_\o)^{\lambda_\o-2} - (\lambda_\o-1)C_M(\lambda_\o) (K_4 M)^{\lambda_\o -2} + \hbox{h.o.,} \label{eq:small_M} 
\end{align}
where the coefficients of the entropy terms $C_a (\lambda_a)$ and $C_M(\lambda_\o)$ are {presented in Appendix C} and the coefficient of the interlayer interaction term $D_m(\lambda_\o)$ of the r.h.s. of Eq.~\ref{eq:small_mo} is also presented in Appendix C. This term needs to be considered, because it is negative and contributes to the first-order transition.

To obtain $T_s$, we first consider the lowest-order terms of Eqs.~\eqref{eq:small_mn} and \eqref{eq:small_mo} and obtain the following: 
	\begin{align}
	\left(1- \dfrac{\langle k_\o^2 \rangle}{\langle k_\o \rangle} \dfrac{1}{T} \right) \left(1- \dfrac{\langle k_\n^2 \rangle}{\langle k_\n \rangle} \dfrac{1}{T} \right) - \dfrac{1}{T^2}  \langle k_\n \rangle \langle k_\o \rangle = 0 \,. \label{eq:combination}
	\end{align}
This equation has two solutions of $T$, denoted as $T_\ell$ and $T_h$.  

It is guaranteed that the l.h.s of Eq.~\eqref{eq:small_M} is positive as $T\to T_h^-$ as long as $x < x_M$ with 
\begin{align}
x_M\equiv T_h{\langle k_\o \rangle}/{\langle k_\o^2 \rangle}.
\label{eq:x_M}
\end{align}
Then $M$ is written within a leading order as 
\begin{align}
M\simeq \frac{D_m(\lambda_\o)}{\langle k_\o \rangle [1- x T_h/(x_M T)]} (K_2 m_\o)^{\lambda_\o -2} + \hbox{h.o.}\,.
\label{eq:M_and_m_relation}
\end{align}
This implies that $O(M) \ll O(m_a)$ near $T_h^-$ for $\lambda_\o >3$. Using this relation \eqref{eq:M_and_m_relation}, we obtain the self-consistency relations for $m_a$ with  leading terms as follows: 
\begin{align}
m_a \langle k_\n \rangle \langle k_\o \rangle \left(1- \frac{T_\ell}{T} \right)\left(1- \frac{T_h}{T} \right) \simeq - (\lambda_a-1) E_{\bar a} C_a (\lambda_a)(K_2 m_a)^{\lambda_a-2} - (\lambda_{\bar a}-1) F_{a\bar a} C_{\bar a}(\lambda_{\bar a})(K_2 m_{\bar a})^{\lambda_{\bar a}-2} \label{eq:eliminating_m}\,,
\end{align}
where
$E_{\bar a} =\langle k_{\bar a} \rangle \left(1- \dfrac{\langle k_{\bar a}^2 \rangle / \langle k_{\bar a} \rangle}{T} \right)$ and $F_{a\bar a} = \dfrac{1}{T} \langle k_\n \rangle \langle k_\o \rangle$.

As $T\to T_h^-$, $m_a \to 0$ and thus $M \to 0$ in Eq.~\eqref{eq:M_and_m_relation}. This implies that the PT from Baxter to Para phase is continuous. We confirm that a second-order transition occurs at $T_h$. This temperature is denoted as a critical temperature $T_s\equiv T_h$. 

Using Eq.~\eqref{eq:M_and_m_relation}, we expand the free energy density of Eq.~\eqref{eq:freeEnergyDensity} with respect to $m_a$ up to the three lowest order terms: 

\begin{itemize}
\item[(i)] For $\lambda_\n > \lambda_\o$,
\begin{align}
f(m_\o) \simeq A_\o  K_2 m_\o ^2 \left(1-\frac{T_s}{T}\right)  + C_\o (\lambda_\o) (K_2 m_\o)^{\lambda_\o - 1} + C_\n (\lambda_\n) (B_\o  m_\o)^{\lambda_\n - 1}-\frac{1}{2}\frac{K_4 [D_m(\lambda_\o)]^2}{\langle k_\o  \rangle [1- x T_s/(x_M T)]} (K_2 m_\o)^{2(\lambda_\o -2)} + \hbox{h.o.}, 
\label{eq:free_energy_x<1_o}
\end{align}
where $A_\o $ and $B_\o$ are functions of $\lambda_\a$ and $K_2$, for which the explicit formula is as follows:
\begin{align}
A_\o  (\lambda_\o) = \dfrac{K_2 \langle k_\n  \rangle^2 K_2 \langle k_\o  \rangle^2}{\langle k_\n  \rangle - K_2 \langle k_\n ^2 \rangle}, \qquad B_\o  (\lambda_\o) 
= \dfrac{K_2 \langle k_\n  \rangle K_2 \langle k_\o  \rangle}{\langle k_\n  \rangle - K_2 \langle k_\n ^2 \rangle}.
\end{align}  
There is an additional negative higher order term as follows:
\begin{align}
\dfrac{D_m(\lambda_\o)}{\langle k_\o \rangle [1- x T_h/(x_M T)]} (K_2 m_\o)^{\lambda_\o} \left[\dfrac{\lambda_{\o}-1}{\lambda_{\o}-3}(k^{\o}_{\textrm{min}})^{3-\lambda_{\o}} + 2 \dfrac{B_\o}{K_2}\langle (k^{\n}) \rangle \langle (k^{\o})^2 \rangle +  (\dfrac{B_\o}{K_2})^2 \langle (k^{\n})^2 \rangle \langle (k^{\o}) \rangle \right] \label{eq:additional_x<1_o}.
\end{align}
Note that as $x \to x_M^-$, the magnitude of Eq.~\eqref{eq:additional_x<1_o} becomes comparable to that of the term with $C_3$.

\item[(ii)] For $\lambda_\n < \lambda_\o $,
\begin{align}
f(m_\n) \simeq 
A_\n K_2 m_\n ^2 \left(1-\frac{T_s}{T} \right) + C_\n (\lambda_\n) (K_2 m_\n)^{\lambda_\n - 1} + C_\o (\lambda_\o) (B_\n  m_\n)^{\lambda_\o - 1} - \frac{1}{2}\frac{K_4 [D_m(\lambda_\o)]^2}{\langle k_\o  \rangle [1- x T_s/(x_M T)]} (B_\n  m_\n)^{2(\lambda_\o -2)} + \hbox{h.o.},
\label{eq:free_energy_x<1_n}
\end{align} 
\end{itemize}
where $A_\n $ and $B_\n$ are functions of $\lambda_\a$ and $K_2$, for which the explicit formula is as follows:
\begin{align}
A_\n  (\lambda_\n)  = \dfrac{K_2 \langle k_\n  \rangle^2  K_2 \langle k_\o  \rangle^2}{\langle k_\o  \rangle - K_2 \langle k_\o ^2 \rangle} \qquad B_\n  (\lambda_\n) = \dfrac{K_2 \langle k_\n  \rangle  K_2 \langle k_\o  \rangle}{\langle k_\o  \rangle - K_2 \langle k_\o^2 \rangle}.
\end{align}
Note that $C_ a  (\lambda_ a  )$ and $D_{m}(\lambda_\o)$ are always positive.

The phase diagram in the space of $[x, T^{-1}]$ depends on $\lambda_\n$ and $\lambda_\o$ as shown in Fig.~\ref{fig:fig4}. These phase diagrams reveal the nature of the observed PTs and  can be obtained by examining the profiles of the free energy density for different $x$ and $T$ values for given $\lambda_\n$ and $\lambda_\o$. To be concrete, here we consider the case of $\lambda_\n=3.53$ and $\lambda_\o=3.90$, for which we obtain the phase diagram similar to that of the original AT model with the {exponent of degree distribution} $\lambda > \lambda_c$.

\subsection{Case $x > x_M$}

The self-consistency relations Eqs.~\eqref{eq:equationOfState_m} and \eqref{eq:equationOfState_M} are expanded in terms of $m_a$ and $M$ as follows: 
\begin{align}
m_\n  \langle k_\n  \rangle \left(1- K_2 \dfrac{\langle k_\n^2 \rangle}{\langle k_\n \rangle} \right)
&\simeq K_2 m_\o \langle k_\n  \rangle \langle k_\o  \rangle 
- (\lambda_\n-1)C_\n (\lambda_\n) (K_2 m_\n)^{\lambda_\n -2} + (K_4 M) (K_2 m_\n) \langle k_\n ^2 \rangle \langle k_\o  \rangle + (K_4 M)  (K_2 m_\o) \langle k_\n  \rangle \langle k_\o ^2 \rangle + \hbox{h.o.}\, \label{eq:small_m_n_>} \\
m_\o  \langle k_\o  \rangle \left(1- K_2 \dfrac{\langle k_\o^2 \rangle}{\langle k_\o \rangle} \right)&\simeq  K_2 m_\n \langle k_\n  \rangle  \langle k_\o  \rangle - (\lambda_\o-1)C_\o (\lambda_\o) (K_2 m_\o)^{\lambda_\o -2} \nonumber \\
&+ D_M(\lambda_\o) (K_4M)^{\lambda_\o -3} (K_2 m_\o) -\left[\int_0^1dk_\o P_d(k_\o) \tanh(K_4 M k_\o) k_\o^2 \right] (K_2 m_\o) \, 
+ (K_4 M) (K_2 m_\n) \langle k_\n  \rangle \langle k_\o ^2 \rangle + \hbox{h.o.}\, \label{eq:small_m_o_>} \\ \cr
M \langle k_\o  \rangle \left(1- x K_2 \dfrac{\langle k_\o^2 \rangle}{\langle k_\o \rangle} \right)
& \simeq -(\lambda_\o-1)C_M(\lambda_d) (K_4 M)^{\lambda_\o -2}
+ (\lambda_\o -3)D_M(\lambda_\o) (K_4 M)^{\lambda_\o -4}(K_2 m_\o)^2 + \hbox{h.o.,}\,  \label{eq:small_M_>} 
 \end{align}
where $D_M(\lambda_\o) >0$ increases monotonically with $\lambda_\o $. This coefficient is explicitly derived in Appendix C. These expansions are valid for $3 < (\lambda_\n, \lambda_\o) < 4$ due to the power of the third term of the r.h.s of Eq.~\eqref{eq:small_m_o_>}.

When $x>x_M$, $x\langle k_\o^2 \rangle/\langle k_\o \rangle >T_h$ and the l.h.s. of Eq~\eqref{eq:small_M_>} becomes negative for $T> T_h$. On the other hand, the first term of the r.h.s. of Eq.~\eqref{eq:small_M_>} is also negative; however, the second term is positive. So, the first term is comparable to the l.h.s., leading to $M \sim (T_{s,M}/T-1)^{1/(\lambda_\o -3)}$, where $T_{s,M} \equiv x{\langle k_\o^2 \rangle}/{\langle k_\o \rangle}$. Thus, $M$ exhibits a continuous transition at $T_{s,M}$, corresponding to the continuous transition curve starting from CE$_2$ in Fig.~\ref{fig:fig3}. Note that this formula is the same as the one of the Ising model on a single SF network~\cite{Doro_book}. For further discussions, $M_*$ is defined as 
$$M_* \equiv \dfrac{1}{K_4}\Big[\dfrac{\langle k_\o \rangle (T_{s,M}/T-1)}{(\lambda_\o-1) K_4 C_M(\lambda_\o)}\Big]^{1/(\lambda_\o -3)}.$$
Next, to determine a critical temperature (denoted as $T_{s,m}$) for $m_a$, we first rewrite Eq.~\eqref{eq:small_M_>} as 
\begin{align}
M\simeq M_* + \frac{D_M(\lambda_\o)}{\langle k_\o  \rangle (T_{s,M}/T-1) } (K_4 M_*)^{\lambda_\o -4} (K_2 m_\o)^{2}+\hbox{h.o}.
\label{eq:M_and_m_relation_>}
\end{align}
We consider the linear terms of $m_a$ in Eqs.~\ref{eq:small_m_n_>} and \ref{eq:small_m_o_>}, and substitute $M$ with $M_*$. Using a similar technique used in Eq.~\eqref{eq:combination}, we obtain the following:
\begin{align}
\langle k_\n  \rangle \langle k_\o  \rangle
\left(1- \dfrac{\langle k_\n^2 \rangle / \langle k_\n \rangle + g_\n (M_*)}{T} \right) \left(1- \dfrac{\langle k_\o^2 \rangle / \langle k_\o \rangle + g_\o (M_*)}{T} \right)  
- \left( \dfrac{\langle k_\n  \rangle \langle k_\o  \rangle }{T} + \dfrac{K_4 M_* \langle k_\n  \rangle \langle k_\o ^2 \rangle}{T} \right)^2 = 0 \,, \label{eq:comb2}
\end{align}
where
\begin{align}
g_\n (M_*) \langle k_\n \rangle = K_4 M_* \langle k_\o  \rangle \langle k_\n ^2 \rangle \,, \quad \textrm{and} \quad g_\o (M_*) \langle k_\o \rangle = D_M(\lambda_\o) (K_4 M_*)^{\lambda_\o -3}-\int_0^1 dk_\o P_d(k_\o) \tanh(K_4 M_* k_\o) k_\o^2 \,. \nonumber
\end{align}
Eq.~\eqref{eq:comb2} has two solutions for $T$, denoted as $T_{\ell}'$ and $T_{h}'$ ($T_{\ell}' < T_{h}'$). Using the relation \eqref{eq:M_and_m_relation_>}, we can obtain a self-consistency relation for $m_a$ within the leading order as follows: 
\begin{align}
m_ a   \langle k_\o \rangle  \langle k_\n  \rangle \left(1- \frac{T_\ell'}{T} \right)\left(1- \frac{T_{h}'}{T} \right) 
\simeq - (\lambda_a-1) E_{\bar a  }'  C_ a  (\lambda_a) \, (K_2 m_ a  )^{\lambda_ a  -2} - (\lambda_{\bar a}-1) F_{a\bar a  }' C_{\bar a  }(\lambda_{\bar a  }) \,(K_2 m_{\bar a  })^{\lambda_{\bar a  }-2} \,,
\end{align}
where
$E_{\bar a}' =\langle k_{\bar a} \rangle \left(1- \dfrac{\langle k_{\bar a}^2 \rangle / \langle k_{\bar a} \rangle + g_{\bar a} (M_*)}{T} \right)$ and $F_{a\bar a  }' = \dfrac{1}{T} \langle k_\o \rangle \langle k_\n  \rangle + \dfrac{1}{T} K_4 M_* \langle k_\o ^2 \rangle \langle k_\n  \rangle$.
We find that near $T_{h}'$, $m_a$ converges to zero continuously, whereas $M$ remains in $\mathcal{O}(1)$. Hence, we regard $T_{h}'$ as the critical temperature $T_{s}'$ of $m_a$. Note that $M$ has the critical temperature $T_{s,M}$ separately, given as $x \langle k_\o^2 \rangle / \langle k_\o  \rangle$, which is higher than $T_{s}'$.

Using Eq.~\eqref{eq:M_and_m_relation_>}, we expand the free energy density of Eq.~\eqref{eq:freeEnergyDensity} with respect to $m_a$ up to the three lowest order terms: 
\begin{itemize}
\item[(i)] For $\lambda_\n > \lambda_\o$,
\begin{align}
f(m_\o) 
&\simeq f_0(M_*) + A_\o' K_2 m_\o^2 \left(1-\frac{T_s'}{T} \right) + C_\o (\lambda_\o) (K_2 m_\o)^{\lambda_\o - 1} + C_\n (\lambda_\n) (B_\o' m_\o)^{\lambda_\n - 1} \cr
& - \frac{K_2[(\lambda_\n-1)C_\n (\lambda_\n)]^2}{\langle k_\n \rangle \left[1 - \left(\langle k_\n^2 \rangle / \langle k_\n \rangle + g_\n(M_*)\right)/T \right]} (B_\o' m_\o)^{2(\lambda_\n -2)} - \frac{1}{2} (\lambda_\o -3)  \dfrac{K_4 \left[ D_M(\lambda_\o) (K_4 M_*)^{\lambda_\o -4}\right]^2}{\langle k_\o  \rangle \left(T_{s,M}/T -1\right)} (K_2 m_\o)^{4} + \hbox{h.o.}\,, 
\label{eq:free_energy_x>1_o}
\end{align}
where $A_\o '$ and $B_\o '$ are functions of $\lambda_a$, $K_2$ and $K_4 M_*$, They are explicitly derived as follows:
\begin{align}
A_\o' (\lambda_\o)
= \dfrac{\left(1 + K_4 M_* \langle k_\o ^2 \rangle / \langle k_\o  \rangle \right)^2 K_2 \langle k_\n  \rangle^2 K_2 \langle k_\o  \rangle^2}{\langle k_\n  \rangle \left[1 - K_2\left(\langle k_\n ^2 \rangle / \langle k_\n \rangle + g_\n(M_*)\right) \right]}, \qquad B_\o' (\lambda_\o) = \dfrac{\left(1 + K_4 M_* \langle k_\o ^2 \rangle / \langle k_\o  \rangle \right) K_2 \langle k_\n  \rangle  K_2 \langle k_\o  \rangle}{\langle k_\n  \rangle \left[1 - K_2 \left(\langle k_\n^2 \rangle / \langle k_\n \rangle + g_\n(M_*)\right) \right]}.
\end{align}
There is an additional negative higher order term as follows:
\begin{align}
\dfrac{D_M(\lambda_\o)}{\langle k_\o  \rangle (T_{s,M}/T-1) } (K_4 M_*)^{\lambda_\o -4} (K_2 m_\o)^{4} \left[ \dfrac{\lambda_{\o}-1}{\lambda_{\o}-3}(k^{\o}_{\textrm{min}})^{3-\lambda_{\o}} + 2 \dfrac{B_\o'}{K_2}\langle (k^{\n}) \rangle \langle (k^{\o})^2 \rangle +  (\dfrac{B_\o'}{K_2})^2 \langle (k^{\n})^2 \rangle \langle (k^{\o}) \rangle \right] \label{eq:additional_x>1_o}
\end{align}
Note that as $x \to x_M^+$, Eq.~\eqref{eq:additional_x<1_o} becomes close in its magnitude to the term with $C_3$ and thus, play a similar role to the term with $C_3$ near $x_M^+$.

\item[(ii)] For $\lambda_\n < \lambda_\o $,
\begin{align}
f(m_\n) 
& \simeq f_0(M_*) + A_\n' K_2 m_\n ^2 \left(1-\frac{T_s'}{T} \right)  + C_\n (\lambda_\n) (K_2 m_\n)^{\lambda_\n - 1} + C_\o (\lambda_\o) (B_\n ' m_\n)^{\lambda_\o - 1} \cr
& - \frac{K_2[(\lambda_\o-1)C_\o (\lambda_\n)]^2}{\langle k_\o  \rangle \left[1 - \left(\langle k_\o^2 \rangle / \langle k_\o \rangle + g_\o(M_*)\right)/T \right]} (B_\n' m_\n)^{2(\lambda_\o -2)} - \frac{1}{2} (\lambda_\o -3)  \dfrac{K_4 \left[ D_M(\lambda_\o) (K_4 M_*)^{\lambda_\o -4}\right]^2}{\langle k_\o  \rangle \left(T_{s,M}/T -1\right)} (B_\n' m_\n)^4 + \hbox{h.o.}\,,
\label{eq:free_energy_x>1_n}
\end{align}
where $A_\n'$ and $B_\n'$ are functions of $\lambda_a$, $K_2$ and $K_4 M_*$. They are explicitly derived as follows:
\begin{align}
A_\n' (\lambda_\o) = \dfrac{\left(1 + K_4 M_* \langle k_\o^2 \rangle / \langle k_\o  \rangle \right)^2 K_2 \langle k_\n  \rangle^2  K_2 \langle k_\o  \rangle^2}{\langle k_\o  \rangle \left[1 - K_2 \left(\langle k_\o ^2 \rangle / \langle k_\o \rangle + g_\o(M_*)\right) \right]}, \qquad B_\n' (\lambda_\o) = \dfrac{\left(1 + K_4 M_* \langle k_\o^2  \rangle / \langle k_\o  \rangle \right)  K_2 \langle k_\n  \rangle  K_2 \langle k_\o  \rangle}{\langle k_\o  \rangle \left[1 - K_2\left(\langle k_\o ^2 \rangle / \langle k_\o \rangle + g_\o(M_*)\right) \right]}.
\end{align}Here, first two $C_ a$ terms are positive, like the case $x<x_M$,
\end{itemize}

The first two $C_a$ terms in Eqs.\eqref{eq:free_energy_x>1_o} and \eqref{eq:free_energy_x>1_n} are positive, as for the case $x < x_M$, whereas the next two terms containing $C_a$ and $D_M$ are negative. The $2(\lambda_a-2)$-order terms with $C_a$ are finite, whereas the $D_M$ term diverges as $T\to T_{s,M}$. Thus, the $D_M$ term contributes to the formation of a global minimum of $f(m_a)$ as $T$ is decreased $T_f$ and $x \to x_M^+$.

\subsection{Case $x \approx x_M$}

\subsubsection{Case $\lambda_{\rm n} \ge  \lambda_{\rm o}$}

In this case, $O(m_a) \sim O(M)$ near $T_s$ and the free energy density of  Eq.~\eqref{eq:freeEnergyDensity} is expanded with respect to $m_\o$ as follows:
\begin{align}
f(m_\o)
	& \simeq A_\o  K_2 m_\o ^2 \left(1-\frac{T_s}{T} \right) + \frac{1}{2} K_4 M^2 \langle k_\o  \rangle \left(1 - \frac{T_s}{T}\right) + C_\o(\lambda_\o , r_\o) (K_2 m_\o)^{\lambda_\o - 1} + C_\n (\lambda_\n) (B_\o m_\o)^{\lambda_\n - 1} + C_3(\lambda_\n ,\lambda_\o , r_0) (K_2 m_\o)^3 + \hbox{h.o.}\,, \label{eq:free_energy_x=1_II}
\end{align}
where $A_\o $ and $B_\o$ are functions of $\lambda_a$ and $K_2$ that are explicitly derived in the $x<x_M$ case. $C_\o(\lambda_\o ,r_0)$ with $r_0 \equiv K_4 M/K_2 m_\o $ is $\mathcal{O}(1)$. Explicit formulas of the coefficients are given in Appendix C.
There is another negative term with $C_3$, which is defined as follows:
\begin{align}
C_3(r_0, \lambda_\o, \lambda_\n) =  r_0 \left[ \dfrac{\lambda_{\o}-1}{\lambda_{\o}-3}(k^{\o}_{\textrm{min}})^{3-\lambda_{\o}} + 2 \dfrac{B_\o}{K_2}\langle (k^{\n}) \rangle \langle (k^{\o})^2 \rangle +  (\dfrac{B_\o}{K_2})^2 \langle (k^{\n})^2 \rangle \langle (k^{\o}) \rangle \right]. \label{eq:additional_x=1}
\end{align}
Note that this term does not appear in the original AT model defined on SF network.

We note that the $C_\o$ term is a leading order term at $T_s$, and $C_\o$ decreases monotonically with $\lambda_\o$. Thus the sign of $C_\o$ can change depending on the magnitude of $\lambda_\o$. This feature does not appear for both $0 < x < x_M$ and $x > x_M$ cases. However, it occurs when $x=x_M$. Numerically $C_\o=0$ at $\lambda_\o \approx 3.503$, equivalent to $\lambda_c$ introduced earlier in Secs. II and III: $C_\o$ becomes positive for $\lambda_\o < \lambda_c$, whereas it is negative for $\lambda_\o > \lambda_c$. On the other hand, $C_\n$ and $C_3$ are always positive and negative, respectively.

\subsubsection{Case $\lambda_{\rm n} = \lambda_{\rm o}$ }

When $\lambda_\n = \lambda_\o$, the $C_\o$ and $C_\n$ terms in Eq.~\eqref{eq:free_energy_x=1_II} are of the same order. Thus, the two terms are combined and denoted as $C_\o'(\lambda_\o, r_0) (K_2 m_\o)^{\lambda_\o-1}$. The sign of $C_\o'$ depends on $\lambda_\o$, equivalently $\lambda_\n$. Numerically $C_\o'$ can be zero at a certain $\lambda_\o$, denoted as $\lambda_e$, estimated to be $ \approx 3.605$. $C_\o'$ becomes positive for $\lambda_\o < \lambda_e$ and negative otherwise.

Similar to the previous case $\lambda_\n > \lambda_\o$, a discontinuous transition always occurs for $C_\o' < 0$. However, depending on relative magnitude between two terms $C_\o'$ and $C_3$, either a discontinuous or continuous transition occurs for $C_\o' > 0$. Note that successive discontinuous transitions do not occur for $\lambda_\n = \lambda_\o$.

\subsubsection{Case $\lambda_{\rm n} < \lambda_{\rm o}$ }

When $\lambda_\n < \lambda_\o$, $\mathcal{O}(m_a) \ll \mathcal{O}(M)$, the free energy density of Eq.~\eqref{eq:freeEnergyDensity} is expanded with respect to $m_\n$ as follows:

\begin{align}
f(m_\n) 
& \simeq f_0(M_*) + A_\n' K_2 m_\n ^2 \left(1-\frac{T_s}{T} \right)  + C_\n (\lambda_\n) (K_2 m_\n)^{\lambda_\n - 1} + C_\o (\lambda_\o) (B_\n' m_\n)^{\lambda_\o - 1} \cr
& - \frac{K_2[(\lambda_\o-1)C_\o (\lambda_\n)]^2}{\langle k_\o \rangle \left[1 - \left(\langle k_\o^2 \rangle / \langle k_\o \rangle + g_\o(M_*)\right)/T \right]} (B_\n' m_\n)^{2(\lambda_\o -2)} - \frac{1}{2} (\lambda_\o -3)  \frac{ K_4 \left[ D_M(\lambda_\o) (K_4 M_*)^{\lambda_\o -4}\right]^2}{\langle k_\o  \rangle \left(T_{s,M}/T -1\right)} (B_\n' m_\n)^4 + \hbox{h.o.}. \nonumber\,
\end{align}
This formula is exactly the same as Eq.~\eqref{eq:free_energy_x>1_n}, derived in $x>x_M$ case for continuous transitions.

\section{The susceptibility near the critical temperature}

\subsection{$m_a$ - magnetization}

\subsubsection{for $x \le x_M$}
Now, we consider the susceptibility at the critical temperature $T_s$ for weak interlayer interaction $x<x_M$ case. We can omit the higher order terms in $m_a$ and $M$ when $m_a$ and $M$ are very small, we expand the self-consistency relations for $m_a$ \eqref{eq:equationOfState_m_H} with respect to $m_a$ and $M$ as follows:
	\begin{align}
	m_a   \langle k_ a   \rangle
	&\simeq\Omega_a \langle k_ a  ^2 \rangle + \Omega_{\bar a  } \langle k_\o  \rangle \langle k_\n  \rangle - (\lambda_a-1)C_ a  (\lambda_a)(\Omega_a)^{\lambda_ a  -2} \,. \label{eq:equationOfState_m_H_expand}
	\end{align}
To obtain critical exponent  $\gamma$ for each $m_a$-magnetization, repectively, we consider the lowest-order terms of the self-consistency relations Eqs.~\eqref{eq:equationOfState_m_H_expand} and then we obtain the following: 
	\begin{align}
	m_ a \langle k_\n  \rangle \langle k_\o  \rangle \left(1 - \dfrac{T_\ell}{T}\right)\left(1 - \dfrac{T_s}{T}\right) 
	& \simeq E_{\bar a} \left(H_a \langle k_a  ^2 \rangle + H_{\bar a} \langle k_\o \rangle \langle k_\n  \rangle - (\lambda_a-1) C_a (\lambda_a) (\Omega_a)^{\lambda_a-2}\right) \cr 
	&+ F_{a \bar a} \left(H_{\bar a} \langle k_{\bar a}^2 \rangle + H_a \langle k_a \rangle \langle k_{\bar a} \rangle - (\lambda_{\bar a}-1) C_{\bar a}(\lambda_{\bar a}) (\Omega_{\bar a})^{\lambda_{\bar a}-2}\right) + \hbox{h.o.}\,, \label{eq:equationOfState_m_H_expand_expand}
	\end{align}
where $E_{\bar a} = \langle k_{\bar a} \rangle \left(1- \dfrac{\langle k_{\bar a}^2 \rangle / \langle k_{\bar a} \rangle}{T} \right)$ and $F_{a \bar a} = \dfrac{1}{T} \langle k_\o  \rangle \langle k_\n  \rangle$.

In order to derive the susceptilbility from magnetization, we take partial derivative with respect to $H_a$ and then take the $H_a$ and $H_4 \rightarrow 0$ limit. We have two equations for the susceptibility as follows:
	\begin{align}
	\chi_a   \langle k_a \rangle \langle k_{\bar a  } \rangle (1 - \dfrac{T_\ell}{T})(1 - \dfrac{T_s}{T}) 
	& \simeq E_{\bar a} \left( \langle k_a  ^2 \rangle - (\lambda_a-1)(\lambda_ a-2) C_a(\lambda_a) (K_2 m_a)^{\lambda_ a  -3} K_2 \chi_a   \right) \cr
	&+ F \left( \langle k_a \rangle \langle k_{\bar a} \rangle - (\lambda_{\bar a}-1)(\lambda_{\bar a}-2)C_{\bar a}(\lambda_{\bar a}) (K_2 m_{\bar a  })^{\lambda_{\bar a}-3} K_2 \dfrac{\partial m_{\bar a  }}{\partial H_2^ a  } \right) + \hbox{h.o.}
	\label{eq:susceptibility_m} \,.
	\end{align}
When $\lambda_\o  < \lambda_\n $, the susceptibility near the critical temperature $T_s^-$  is written as follows:	
\begin{align}
	\chi_\o  & \langle k_\n  \rangle \langle k_\o  \rangle (1 - \dfrac{T_\ell}{T})(1 - \dfrac{T_s}{T}) 
	\approx E_\n  \langle k_\o ^2 \rangle + F \langle k_\o  \rangle \langle k_\n  \rangle + (\lambda_\o -2) \langle k_\o \rangle \langle k_\n \rangle (1 - \dfrac{T_\ell}{T})(1 - \dfrac{T_s}{T}) \chi_\o  + \hbox{h.o.} \,, \cr
	\chi_\n  & \langle k_\n  \rangle \langle k_\o  \rangle (1 - \dfrac{T_\ell}{T})(1 - \dfrac{T_s}{T})
	\approx E_\o \langle k_\n ^2 \rangle + F \langle k_\o  \rangle \langle k_\n \rangle + (\lambda_\o -2) \langle k_\o \rangle \langle k_\n \rangle (1 - \dfrac{T_\ell}{T}) (1 - \dfrac{T_s}{T}) \dfrac{E_\o}{E_\n} \dfrac{\partial m_\o }{\partial H_\n } + \hbox{h.o.} \,.
	\end{align}
To obtain the susceptibility near $T_s^-$, we use the following relation
	\begin{align}
	E_\n \left(-(\lambda_\o-1) K_2 C_\o  (\lambda_\o) (K_2 m_\o)^{\lambda_\o -3} \right) \approx {\langle k_\o  \rangle \langle k_\n  \rangle} (1-\dfrac{T_\ell}{T})(1-\dfrac{T_s}{T}) &~~\hbox{ for }~~ T \to T_s^-.
	\end{align}
To get $\chi_\n $, we need to compute the $\partial m_\o /\partial H_\n $ term.
The partial derivative of $m_\o $ in terms of $H_\n $ is given as follows:
\begin{align}
\dfrac{\partial m_\o}{\partial H_\n} \langle k_\o  \rangle \langle k_\n  \rangle (1 - \dfrac{T_\ell}{T})(1 - \dfrac{T_s}{T}) 
& \simeq E_\n  \left( \langle k_\o  \rangle \langle k_\n  \rangle + (\lambda_\o-1)(\lambda_\o-2) C_\o (\lambda_\o) (K_2 m_\o)^{\lambda_\o-3} K_2 \dfrac{\partial m_\o}{\partial H_\n}\right) + F \langle k_\n ^2 \rangle + \hbox{h.o.} \cr
& \simeq E_\n  \langle k_\o  \rangle \langle k_\n  \rangle + F  \langle k_\n ^2 \rangle + (\lambda_\o-2) \langle k_\o  \rangle \langle k_\n  \rangle (1 - \dfrac{T_\ell}{T}) (1 - \dfrac{T_s}{T}) \dfrac{\partial m_\o }{\partial H_\n} + \hbox{h.o.} 
\label{eq:susceptibility_m_n} \,.
\end{align}
From Eq.~\eqref{eq:susceptibility_m_n}, we obtain that $\partial m_\o / \partial H_\n  \approx (T_s - T)^{-1}$ for $T \to T_s^-$.
Using this, the susceptibility of $m_a$ is obtained as 
	\begin{align}
	\chi_\o  \approx \left\{ \begin{array}{cc}
	(T - T_s)^{-1} &~~\textrm{for}~ T_s^+, \\ \\
	(T_s - T)^{-1} &~~\textrm{for}~ T_s^-,
	\end{array} \right. \quad\quad \hbox{and} \quad\quad
	\chi_\n  \approx \left\{ \begin{array}{cc}
	(T - T_s)^{-1} &~~\textrm{for}~ T_s^+, \\ \\
	(T_s - T)^{-1} &~~\textrm{for}~ T_s^-.
	\end{array} \right.
	\label{eq:susceptibility_m_<} \,
	\end{align}
Here, we take the limit $m_\o \to 0$ for near $T_s^+$ to Eq. \eqref{eq:susceptibility_m}. For the $\lambda_\n <\lambda_\o$ case, the susceptibility of $m_a$ is obtained by similar computation process same as for the $\lambda_\o <\lambda_\n$ case.
Thus, the critical exponent of $\gamma_{m\pm}$ of magnetization is always $1$ for all cases. Then, the scaling relation, $a+2\beta_m+\gamma_{m-}=2$, is satisfied for each $m_a$-magnetization\cnew{,} respectively. 

Now, we compute the susceptibility at CE point as boundary point of continuous PT regime. 
Since {the location of magnetization jumps to a certain finite} at the CE, the magnitude of the magnetization is much greater than 0, the {perturbative} expansions with respect to $m,M$ is not valid any longer at CE point. Thus, we should keep the integral formula written in self-consistent relation Eqs.~\eqref{eq:equationOfState_m_H} and \eqref{eq:equationOfState_M_H} as follows. In order to obtain the susceptibility for $m_a$ magnetization, we take a partial derivative of self-consistent relation for $m_a$ \eqref{eq:equationOfState_m_H} with respect ot $H_2$ and take $H_2,H_4 \rightarrow 0$ limit, then the susceptibility is written as follows:
	\begin{align}
	\chi_\o  = \dfrac{\A_{\o\o} + \A_{\o\n} K_2 \partial m_\n /\partial H_\o + \A_{\o M} K_4 \partial M/\partial H_\o}{\langle k_\o \rangle - K_2 \A_{\o\o}} \,.
	\label{chi_m_II_explicit}
	\end{align}
To evaluate Eq.~(\ref{chi_m_II_explicit}), we also should compute the 
	\begin{align}
	\left. \dfrac{\partial m_\n }{\partial H_\o } \right|_{H_a,H_4\rightarrow 0} \textrm{and } \left. \dfrac{\partial M}{\partial H_\o } \right|_{H_a,H_4\rightarrow 0}
	\end{align} terms. Thus, we firstly take a derivative of self-consistent relation for $m_\n $ and $M$ \eqref{eq:equationOfState_m_H},\eqref{eq:equationOfState_M_H} with respect to $H_\o$ and then take the limit $H_a$ and $H_4 \rightarrow 0$, we obtain as follows:
	\begin{align}
	\dfrac{\partial m_\n}{\partial H_\o} = \dfrac{\A_{\n\o} + \A_{\n\o} K_2 \chi_\o + \A_{\n M} K_4 \partial M/\partial H_\o }{\langle k_\n  \rangle - K_2 \A_{\n\n}}, \quad
	\dfrac{\partial M}{\partial H_\o } = \dfrac{\A_{M \o} + \A_{M \o} K_2 \chi_\o  +  \A_{M \n} K_2 \partial m_\n /\partial H_\o }{\langle k_\o  		\rangle - K_4 \A_{MM}}.
	\label{chi_m_II_explicit_partial}
	\end{align}
At the CE point, $\chi_\o$ is computed similarly to Eq.~(\ref{eq:susceptibility_m}) at $T_s^+$, where $m_a = M = 0$. For $T_s^-$, $\chi_m$ can be {obtained numerically from} Eqs.~(\ref{chi_m_II_explicit}), (\ref{chi_m_II_explicit_partial}). We can confirm that the susceptibility has a certain finite value at $T_s^-$ by numerical computations.

\subsubsection{Case $x>x_M$}
Otherwise, for $x>x_M$, we expand the self-consistency relations for $m_a$ with respect to $m_a$ and $M$ as follows:
	\begin{align}
	m_ a   \langle k_\n  \rangle \langle k_\o  \rangle (1 - \dfrac{T_\ell'}{T})(1 - \dfrac{T_s'}{T}) 
	& \simeq E_{\bar a  }' \left[ H_a   \langle k_a^2 \rangle + H_{\bar a} \langle k_\o  \rangle \langle k_\n  \rangle - (\lambda_a - 1) C_ a  (\lambda_a) (\Omega_a)^{\lambda_ a  -2}\right] \cr 
	& + F_{a \bar a}' \left[ H_{\bar a} \langle k_{\bar a}^2 \rangle + H_a   \langle k_a \rangle \langle k_{\bar a} \rangle -  (\lambda_{\bar a} - 1)C_{\bar a}(\lambda_{\bar a}) (\Omega_{\bar a})^{\lambda_{\bar a}-2}\right] + \hbox{h.o.}\,, \label{eq:equationOfState_m_H_expand_expand_>}
	\end{align}
where $E_{\bar a}' =\langle k_{\bar a} \rangle \left(1- \dfrac{\langle k_{\bar a}^2 \rangle / \langle k_{\bar a} \rangle + g_{\bar a} (M_*)}{T} \right)$ and $F_{a\bar a  }' = \dfrac{1}{T} \langle k_\o \rangle \langle k_\n  \rangle + \dfrac{1}{T} K_4 M_* \langle k_\o ^2 \rangle \langle k_\n  \rangle$.

It can be checked easily that Eq.~\eqref{eq:equationOfState_m_H_expand_expand_>} is similar case to Eq.~\eqref{eq:equationOfState_m_H_expand_expand} except the critical temperature $T_s'$ and coefficients $E_{\bar a}'$ and $F'$. Thus, we performed similar calculations as for the $x<x_M$ case considering minor differences between the cases $x<x_M$ and $x>x_M$. By performing similar calculations as for the $x<x_M$ case, we obtain the susceptibility of $m_a$ as follows:
	\begin{align}
	\chi_\o  \approx \left\{ \begin{array}{cc}
	(T - T_s')^{-1} &~~\textrm{for}~ {T>T_s'}, \\ \\
	(T_s' - T)^{-1} &~~\textrm{for}~ {T_s'>T},
	\end{array} \right. \quad\quad \hbox{and} \quad\quad
	\chi_\n  \approx \left\{ \begin{array}{cc}
	(T - T_s')^{-1} &~~\textrm{for}~ {T>T_s'}, \\ \\
	(T_s' - T)^{-1} &~~\textrm{for}~ {T_s'>T}.
	\end{array} \right. 
	\label{eq:susceptibility_m_>} \,
	\end{align}

\subsection{$M$ - magnetization}

Likewise, the self-consistency relation for $M$ \eqref{eq:equationOfState_M_H} can be expanded as 
	\begin{align}
	M \langle k_\o \rangle \simeq \Omega_4 \langle k_\o ^2\rangle - (\lambda_\o -1)C_M(\lambda_\o) \Omega_4^{\lambda_\o -2} - D_m(\lambda_\o) \Omega_{\lambda_\o -2} + \hbox{h.o.}.
	\end{align}
To obtain the susceptibility of $M$, we take partial derivative of the above self-consistency relation with respect to $H_4$ and then taking $H_2$ and $H_4 \rightarrow 0$.
\begin{align}
\chi_M \langle k_\o \rangle \simeq (K_4 \chi_M+1) \langle k_\o ^2\rangle - (\lambda_\o -2) C_M(\lambda_\o) (K_4 \chi_M) (K_4 M)^{\lambda_\o -3} - (\lambda_\o -2) D_m(\lambda_\o) (K_2 m_\o)^{\lambda_\o -3} K_2 \dfrac{\partial m_\o }{\partial H_4} + \hbox{h.o.}\,.
\end{align}
For $x<x_M$, because $M$ is $\mathcal{O}(m_\o ^{\lambda_\o -2})$, $\partial m_\o /\partial H_4$ is very small compared with the $\mathcal{O}(1)$ term. Taking this limit, we obtain the susceptibility of $M$. We also take the limit $m_\o =0$ for $T_s^+$.
\begin{align}
\chi_M \approx 
\left(T - x\dfrac{\langle k_\o^2 \rangle}{\langle k_\o \rangle}\right)^{-1}. 
\label{eq:susceptibility_M_<}
\end{align}
Otherwise, $x \ge x_M$ and $T \to T_{s,M}^{-}$, $M$ can be approximated to $M_{*}$, where $M_{*}$ becomes Ising spin in single SF networks, and $m_a$ is negligible to $M$. Taking this limit, we can obtain the susceptibility of $M$ as follow. We also take the limit $M=0$ for $T_{s,M}^{+}$.
\begin{align}
\chi_M \approx \left\{ \begin{array}{cc}
(T - T_{s,M})^{-1} &~~\textrm{for}~~ T>T_{s,M}, \\ \\
(T_{s,M} - T)^{-1} &~~\textrm{for}~~ T_{s,M}>T.
\end{array} \right. 
\label{eq:susceptibility_M_<}
\end{align}
\end{widetext}

\begin{acknowledgments}
This research was supported by the NRF Grant (No. NRF-2014R1A3A2069005), the KENTECH Research
Grant (KRG2021-01-007) (BK), and the KIAS individual Grants (No. PG064901) (JSL) at Korea Institute for Advanced Study.
\end{acknowledgments}

\end{document}